\documentclass[a4paper,onecolumn,11pt,unpublished]{quantumarticle}
\pdfoutput=1
\usepackage{tabulary}
\usepackage[utf8]{inputenc}
\usepackage[english]{babel}
\usepackage[T1]{fontenc}
\usepackage{amsmath}
\usepackage{hyperref}
\usepackage{fixltx2e}
\usepackage[numbers,sort&compress]{natbib}
\usepackage{xcolor}
\usepackage{enumerate}
\usepackage{xfrac}
\usepackage{graphicx}
\usepackage{amsmath}
\usepackage{amssymb} 
\usepackage{booktabs}
\usepackage{multirow}

\usepackage{microtype}

\begin{document}
\title{Quantum Circuit Optimization Based on Dynamic Grouping and ZX-Calculus for Reducing 2-Qubit Gate Count}

\author{Kai Chen}
\affiliation{School of Computer and Cyber Sciences, Communication University of China, Beijing 100024, China}

\author{Wen Liu*}
\email{lw8206@cuc.edu.cn}
\affiliation{School of Computer and Cyber Sciences, Communication University of China, Beijing 100024, China}
\affiliation{State Key Laboratory of Media Convergence and Communication, Communication University of China, Beijing 100024, China}
\affiliation{Key Laboratory of Convergent Media and Intelligent Technology (Communication University of China), Ministry of Education, Beijing 100034, China}
\author{Guo-Sheng Xu}
\affiliation{School of Cyberspace Security Beijing University of Posts and Telecommunications, Beijing, 100876, China}
\author{Yangzhi Li}
\affiliation{School of Computer and Cyber Sciences, Communication University of China, Beijing 100024, China}
\author{Maoduo Li}
\affiliation{School of Computer and Cyber Sciences, Communication University of China, Beijing 100024, China}
\author{Shouli He}
\affiliation{School of Computer and Cyber Sciences, Communication University of China, Beijing 100024, China}

\maketitle

\begin{abstract}
In the noisy intermediate-scale quantum (NISQ) era, two-qubit gates in quantum circuits are more susceptible to noise than single-qubit gates. Therefore, reducing the number of two-qubit gates is crucial for improving circuit efficiency and reliability. As quantum circuits scale up, the optimization search space becomes increasingly complex, leading to challenges such as low efficiency and suboptimal solutions. To address these issues, this paper proposes a quantum circuit optimization approach based on dynamic grouping and ZX-calculus. First, a random strategy-based dynamic grouping method partitions the circuit into multiple subcircuits. Second, a ZX-calculus guided $k$-step lookahead search performs equivalent subcircuit filtering to minimize two-qubit gate counts. Third, a delay-aware placement method optimizes the recombined circuit to reduce the overall gate count. Finally, simulated annealing iteratively updates the grouping strategy to achieve an optimized two-qubit gate count. Experimental results on benchmark datasets demonstrate the effectiveness and superiority of the proposed method in reducing two-qubit gates. Compared to the original circuits, the approach achieves an average reduction of 18\% in two-qubit gates. It outperforms classical methods with up to 25\% reduction, especially on gf circuits, and shows a 4\% average improvement over heuristic ZX-calculus-based methods, validating its efficiency.
\end{abstract}

\section{Introduction}
Quantum computers exploit the quantum properties of superposition and entanglement to accomplish complex computational tasks that are intractable for classical computers within practical timeframes. For instance, Shor's algorithm \cite{shorAlgorithmsQuantumComputation1994a} transforms the integer factorization problem into one of determining the periodicity of a modular exponential function, thereby reducing the computational complexity from exponential to polynomial order, specifically to $O(n^2 \log n \log\log n)$. Grover's algorithm \cite{groverFastQuantumMechanical1996a} employs quantum parallelism and amplitude amplification to achieve a quadratic speedup for unstructured search problems, reducing the complexity from linear $O(N)$ to $O(\sqrt{N})$. However, the hardware resources required for implementing quantum computation are limited; the more gates a quantum circuit contains, the greater the hardware consumption. Quantum circuits consist of both two-qubit and single-qubit gates, with the former incurring significantly higher implementation costs than the latter. Quantum circuits are also susceptible to noise, which can introduce errors \cite{liHighprecisionPulseCalibration2025} during computation, and two-qubit gates generally exhibit lower fidelity than single-qubit gates, making them more vulnerable to noise-induced errors \cite{wangProtectingQuantumGates2025}. Therefore, to reduce hardware resource costs and error accumulation while improving circuit reliability, it is essential to optimize the number of two-qubit gates in quantum circuits \cite{karuppasamyComprehensiveReviewQuantum2025}.\par

Optimizing the number of two-qubit gates in quantum circuits poses two primary challenges: (1) verifying the functional equivalence between the original and optimized circuits, and (2) identifying equivalent circuits with a reduced number of two-qubit gates. To tackle these challenges, researchers have developed a variety of methods. First, equivalence transformation rules for quantum circuits have been formalized using diverse representations, including the gate-level model \cite{deutschQuantumTheoryChurch1985}, symbolic notations \cite{amyPolynomialtimeTdepthOptimization2014}, directed acyclic graph (DAG) \cite{maslovQuantumCircuitSimplification2008}, and ZX-diagrams \cite{coeckeInteractingQuantumObservables2008}. Based on these formalized models, optimization techniques such as template matching \cite{biswalTemplatebasedTechniqueEfficient2018}, peephole optimization \cite{liuRelaxedPeepholeOptimization2021b}, stochastic search \cite{rosenhahnMonteCarloGraph2023b}, and machine learning \cite{ruizQuantumCircuitOptimization2025}have been employed to discover functionally equivalent circuits with fewer two-qubit gates, thereby improving circuit efficiency. \par

Many researchers utilize rule-based gate optimization methods to reduce the gate counts in quantum circuits. These methods define gate sequences for matching and corresponding replacements that are applied upon successful identification. By predefining equivalence transformation rules, such techniques can generate circuits with equivalent functionality while minimizing gate counts \cite{itenExactPracticalPattern2022b}. In 2018, Nam et al. \cite{namAutomatedOptimizationLarge2018} proposed an efficient pattern-matching algorithm that represents circuits as linear gate sequences to facilitate sequential rule application. The algorithm incorporates strategies such as gate elimination, gate reordering, and phase polynomial merging to simplify circuit structures, focusing on H, RZ, and both single- and two-qubit gates. Additionally, an optimization strategy was devised for two-qubit gates using floating RZ gates, enabling gate reduction under specific structural conditions. In 2022, Bravyi \cite{bravyi6qubitOptimalClifford2022a} introduced a brute-force approach for optimizing 6-qubit Clifford circuits, employing a pruned breadth-first search to construct a database of circuit combinations. The number of CNOT gates serves as a cost metric to identify optimal subcircuits. Xu et al. \cite{xuQuartzSuperoptimizationQuantum2022} introduced the concept of equivalent circuit classes to efficiently encode circuit transformation spaces. By conducting exhaustive searches within these classes, subcircuits with reduced two-qubit gate counts were located and substituted into the original circuit. In 2023, Gao et al. \cite{gaoQuantumCircuitTemplate2023b} applied template matching to optimize mapped circuits. By operating on each circuit block, their method generated connectivity-aware optimized circuits, reducing the use of two-qubit gates. In 2024, Javadi-Abhari et al.  \cite{javadi-abhariQuantumComputingQiskit2024} introduced a multi-layered optimization strategy comprising pattern matching, hardware topology mapping, and gate-level transformations. This approach minimizes two-qubit gate counts, optimizes their ordering, and adapts circuits to hardware constraints, thereby enhancing execution efficiency on NISQ devices while mitigating noise-induced errors. Despite their effectiveness, rule-based optimization approaches heavily rely on predefined transformation rules, which may not accommodate unconventional gate configurations lacking applicable transformations.\par

In addition to rule-based methods, researchers have developed various intermediate representation (IR)-based techniques for quantum circuit optimization. These methods search for equivalent circuit structures through IRs to facilitate optimization. An IR exploits the algebraic properties of quantum circuits by mapping them into alternative representations, where optimization is either performed directly or employed to uncover broader circuit isomorphisms. These candidate transformations are subsequently filtered to obtain circuits with reduced gate counts. Common IR formats include symbolic expressions, directed acyclic graphs (DAGs), and ZX-diagrams. In 2021, Bravyi et al. \cite{bravyiCliffordCircuitOptimization2021b} proposed a symbolic peephole optimization method that projects circuits onto subsets of qubits and replaces entangling CNOT gates with symbolic Pauli operations. Dynamic programming is subsequently used to reorder gate sequences and reduce the overall CNOT count. Also in 2021, Sivarajah et al. \cite{sivarajahT|ketRetargetableCompiler2021} utilized t\texttt{\textbar ket$\rangle$} to optimize noisy quantum circuits, where its intermediate representation is represented as a DAG with additional structural annotations. Their approach employs a peephole optimization strategy to traverse the circuit and identify extended instruction subgraphs. By combining template matching and dynamic programming algorithms, specific subcircuit patterns are detected and replaced with equivalent subcircuits with fewer gates or reduced depth, thereby reducing two-qubit gate usage. Hietala et al. \cite{hietalaVerifiedOptimizerQuantum2021} developed the VOQC framework, which translates circuits into a formally verifiable Simple Quantum Intermediate Representation (SQIR). Optimization within VOQC involves propagating and removing redundant operations, with a focus on reducing two-qubit gates. In 2022, Chen et al. \cite{chenPatternTreeEnhancing2024} proposed a pattern tree framework that organizes transformation rules in a hierarchical structure, incorporating prefix and compacted trees to minimize redundancy. A depth-first scheduling algorithm is integrated with preprocessing to enhance gate reduction efficacy. In 2023, Amanda Xu \cite{xuSynthesizingQuantumCircuitOptimizers2023b} introduced the Queso method, which defines symbolic rewrite rules and synthesizes multiple rule sets. These rules are applied in an iterative fashion to reduce two-qubit gate usage. In 2024, Beaudoin et al. \cite{beaudoinAltGraphRedesigningQuantum2024c} proposed a search-based technique that generates optimized directed acyclic graphs (DAGs) through generative models such as DAG variational autoencoders, GRUs, GCNs, and deep graph generative models. While this method effectively reconstructs optimized circuits from these representations, its scalability is limited to circuits with at most six qubits and thirty-two gates, rendering it impractical for larger-scale systems. Although symbolic and DAG-based rewriting approaches enable circuit optimization via local gate transformations, their reliance on localized rules constrains their capacity to identify equivalent transformations that require long-range entanglement and nonlocal gate commutation.\par

Researchers have extensively investigated ZX-diagram representations as a means of optimizing quantum circuits by exploiting global structural properties of the underlying graphs. ZX-calculus provides a formal framework for translating quantum circuits into graphical representations, wherein circuits are rewritten as ZX-diagrams and subsequently transformed through equivalence-preserving rewrite rules. The theoretical foundation of ZX-calculus was established by Coecke and Duncan in 2011 \cite{coeckeInteractingQuantumObservables2011a}. In 2014, ZX-calculus underwent substantial theoretical developments, particularly in the completeness property and tooling infrastructure. Backens \cite{backensZXcalculusCompleteStabilizer2014} proved the completeness of ZX-calculus for the Clifford gate set, thereby laying the groundwork for rigorous equivalence verification. In 2018, Ng et al. \cite{jeandelCompleteAxiomatisationZXcalculus2018} extended this result to the Clifford+T gate set, establishing its universality for practical quantum computing. In 2020, van de Wetering \cite{vandeweteringZXcalculusWorkingQuantum2020a} formalized the completeness theorem, demonstrating that ZX-diagrams can, in principle, express all forms of quantum computational reasoning. That same year, Kissinger introduced the PyZX framework \cite{kissingerPyZXLargeScale2020}, which provides features including circuit simplification and equivalence checking by rewriting circuits into ZX-diagrams, applying equivalence-preserving ZX rewrite rules, and extracting optimized circuits. Furthermore, Kissinger \cite{kissingerReducingTcountZXcalculus2020a} proposed the use of phase gadgets and phase teleportation within ZX-diagrams to reduce the T-count in quantum circuits. However, due to the instability inherent in circuit extraction from ZX-diagrams, these methods were largely effective only for small-scale circuits and frequently encountered failure when applied to larger circuits with increased qubit counts. In 2023, Staudacher et al. \cite{staudacherReducing2QuBitGate2023a} proposed a two-qubit gate optimization framework that employs heuristic-driven cost estimation within ZX-diagrams. Their method employs stochastic and greedy search strategies, along with node non-fusion constraints and edge simplification heuristics, to identify equivalent circuits with significantly fewer two-qubit gates. In 2024, Nägele et al. \cite{nägeleOptimizingZXdiagramsDeep2024} demonstrated that a reinforcement learning agent, guided by a graph neural network (GNN), can effectively apply ZX rewrite rules and outperform traditional heuristics in optimizing larger ZX-diagrams. In 2025, Jordi Riu \cite{riuReinforcementLearningBased2025a} further advanced this direction by proposing a reinforcement learning-based method utilizing graph-theoretic simplification rules and Proximal Policy Optimization (PPO). GNNs are employed to approximate the policy and value functions, enabling the agent to generate structurally equivalent ZX-diagrams in graphical form. By explicitly accounting for node connectivity and its impact on gate count during circuit extraction, this method achieves measurable reductions in the number of two-qubit gates. Despite these advances, current reinforcement learning approaches suffer from large and unstructured rewrite rule spaces, making the search process inefficient and prone to convergence issues. Moreover, the high computational cost of training, due to the complexity of the action space and extensive environment interaction, limits practical scalability. More broadly, a common limitation across ZX-calculus-based methods is their discrete, single-step application of rewrite rules, which often cannot effectively capture compound transformations necessary for discovering globally optimal circuit representations. \par

Furthermore, researchers have adopted a preprocessing approach based on subcircuit grouping for quantum circuit optimization. This method involves first designing a grouping strategy aligned with the circuit’s optimization objectives, followed by the independent optimization of each subcircuit group. Finally, the optimized subcircuits are reassembled to reconstruct a circuit exhibiting a reduced total gate count. In 2021, Patel et al. \cite{patelRobustResourceEfficientQuantum2021a} proposed partitioning circuits based on qubit count and applying approximate synthesis techniques to reduce circuit depth and gate complexity. Their partitioning strategy considered both structural and functional aspects of the circuit, aiming to form moderately sized blocks. For complex multi-qubit circuits, the methodology decomposed them into blocks with reduced qubit counts based on qubit connectivity and operation sequencing. Multiple approximately equivalent variants were generated for each block, and the optimal one—in terms of CNOT gate minimization—was selected. Also in 2021, Wu et al. \cite{wuQGoScalableQuantum2022b} introduced a partitioning strategy within the Qgo framework, which decomposes circuits into logically independent subblocks using the directed acyclic graph (DAG) representation and qubit layout. Each subblock is subjected to local gate rewriting \cite{davisOptimalTopologyAware2020} for optimization, before reintegrating the blocks to form a globally optimized circuit. In 2022, Weiden et al. \cite{weidenWideQuantumCircuit2022b} proposed a partitioning algorithm that segments circuits into subcircuits constrained to a maximum width of $k$. Each subcircuit is individually optimized and subsequently merged to reduce the total number of two-qubit gates. Sutcliffe \cite{sutcliffeSmarterKPartitioningZXdiagrams2024a} proposed a simulation strategy that improves the efficiency of classical quantum circuit simulation by leveraging optimized $k$-partitioning of ZX-diagrams. This method decomposes a ZX-diagram into $k$ strategically selected subdiagrams, each of which is reduced independently before being reassembled to recover the full circuit amplitude. This partitioning approach leverages the structural properties of ZX representations to minimize computational overhead during simulation. However, these partitioning-driven optimization techniques primarily rely on deterministic strategies. Due to their dependence on fixed transformation rules, such methods often converge to local optima and cannot consistently discover globally optimal circuit configurations.\par

Based on previous research, this work addresses the limitations of existing ZX-calculus-based quantum circuit optimization methods, particularly their restricted search capabilities and their inability—due to limitations in ZX-to-circuit extraction—to consistently identify circuit structures with minimal two-qubit gate counts. To overcome these issues, we propose a dynamic partitioning and ZX-based optimization framework. The core idea is to dynamically divide the circuit into subcircuits and apply a lookahead search strategy within each group. This approach breaks the locality constraint of conventional ZX optimization by exploring a significantly expanded transformation space, enabling the discovery of equivalent circuits with fewer two-qubit gates.

The main contributions of this paper are summarized as follows:
\begin{enumerate}[(1)]
	\item We propose a stochastic strategy to partition quantum circuits into subcircuits based on gate execution order. This method diversifies the search space by covering more potential equivalence classes, thereby increasing the likelihood of discovering optimal subcircuit decompositions.
	
	\item A novel subcircuit optimization method is introduced, which converts subcircuits into ZX diagrams and performs $k$-step lookahead rule matching. This strategy enables effective pruning of suboptimal transformation paths and extraction of equivalent subcircuits with reduced two-qubit gate counts.
	\item We use a global optimization strategy based on delayed gate placement. After optimizing subcircuits individually, the merged circuit is refined using rule-based post-processing to eliminate redundant gates, especially H gates, improving the global gate efficiency.
	\item A simulated annealing algorithm is employed to iteratively update circuit partitioning strategies. Guided by the Metropolis acceptance criterion, this mechanism avoids local optima and effectively converges toward global minima in two-qubit gate count.
	\item Experiments on benchmark circuits show that our approach reduces the average number of two-qubit gates by 18\%. Compared to VOQC, Qiskit, and Quartz, our method achieves superior reductions in two-qubit gates. Against the rule-based Nam optimizer, our approach performs better on gf circuits. Compared to heuristic ZX-based methods, our method reduces the two-qubit gate count by an average of 4\%, and achieves up to 15\% improvement over Nam in cases like adder\_8, demonstrating the effectiveness and robustness of the proposed framework.
	
\end{enumerate}
The remainder of this paper is organized as follows: Section 2 provides background knowledge on quantum computing and the ZX calculus. Section 3 formalizes the quantum circuit optimization problem and presents the proposed dynamic grouping and ZX-calculus-based optimization framework in detail. Section 4 describes the experimental setup and presents a comparative analysis of the results. Section 5 concludes the paper and outlines directions for future research.

\section{Preliminaries and Definitions}

\subsection{Quantum Basics and Circuit Optimization}

\subsubsection{Fundamentals of Quantum Computing}
Quantum computing is a computational paradigm that exploits the principles of quantum mechanics to manipulate quantum information units for efficient parallel computation. It employs qubits as the fundamental units of computation and leverages phenomena such as quantum superposition and entanglement to achieve higher efficiency and potential than classical computers in solving certain problems. Quantum gates serve as the basic operational units, enabling transformations and entanglement of qubit states, thus facilitating the implementation of various quantum algorithms.

Qubits are the fundamental units of quantum information. Unlike classical bits, qubits can exist simultaneously in superpositions of the basis states \(\lvert 0 \rangle\) and \(\lvert 1 \rangle\), or any arbitrary linear combination thereof. The state of a qubit can be represented as \(\lvert \psi \rangle = \alpha \lvert 0 \rangle + \beta \lvert 1 \rangle\), where \(\alpha\) and \(\beta\) are complex coefficients satisfying the normalization condition \(|\alpha|^2 + |\beta|^2 = 1\). This condition ensures that the total probability of the qubit being measured in any state sums to one.

Operations on qubits in quantum computing are realized via quantum gates. Quantum gates are fundamental building blocks that manipulate the states of qubits to implement various quantum computational tasks. A quantum circuit serves as the basic framework of quantum computation, comprising a sequence of quantum gates applied to qubits to realize specific quantum algorithms.

Any quantum circuit can be constructed from a universal gate set, such as the Clifford+T set, which includes Pauli gates, the Hadamard (H) gate, the phase (S) gate, the controlled-NOT (CNOT) gate, and the T gate. The Pauli gates consist of Pauli-X, Pauli-Y, and Pauli-Z, which can be interpreted as 180-degree rotations around the corresponding axes on the Bloch sphere. The Hadamard gate (H) transforms a qubit from a basis state into a superposition state. The CNOT gate entangles two qubits by flipping the state of the target qubit conditional on the control qubit. The phase gate (S) applies a \(\pi/2\) phase shift to the qubit. The T gate, also known as the \(\pi/8\) gate, applies a \(\pi/4\) phase shift. Any single-qubit gate can be approximated by a combination of H, S, and T gates, while any multi-qubit gate can be approximated using CNOT gates combined with arbitrary single-qubit gates.

\subsubsection{Quantum Circuit Optimization}

The efficiency of a quantum circuit directly determines the execution speed and resource consumption of quantum computation. A greater number of gates results in slower execution and higher implementation costs. In quantum computing, each quantum gate introduces a certain amount of error and noise. Excessive gate operations accumulate these errors, thereby degrading the accuracy of computational results. Therefore, logical quantum circuit optimization aims to reduce the number of quantum gates. Pattern matching techniques can be employed to identify reducible substructures within the circuit.

\begin{figure}[h]
	\centering
	\includegraphics[width=90mm,height=20mm]{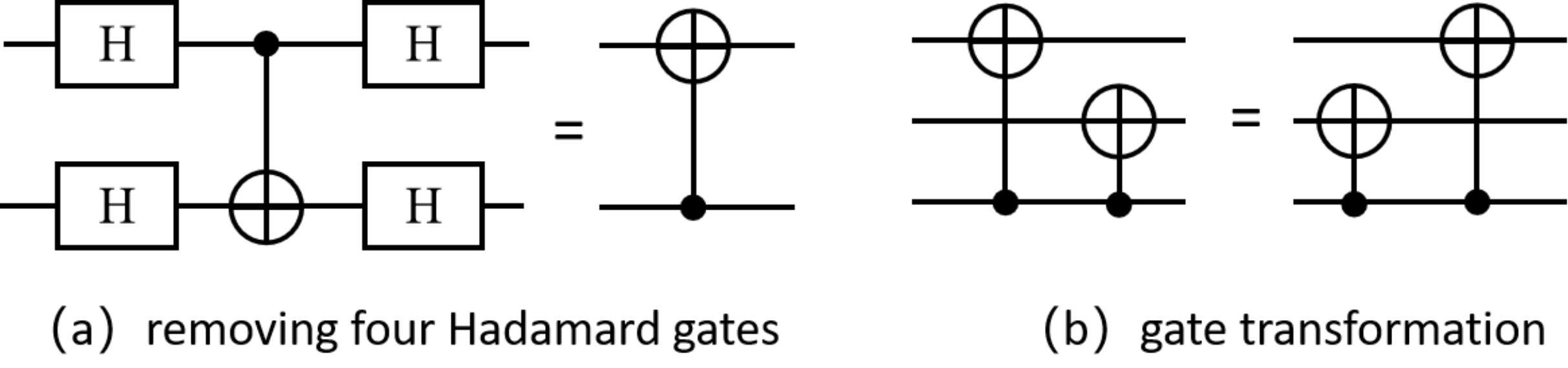}
	\caption{gate transformation}
	\label{fig:figure1}
\end{figure}

A quantum circuit in quantum computing consists of a sequence of quantum gates applied to qubits. Using equivalence rules, quantum gate transformations can be performed, such as gate cancellation and gate commutation. Gate cancellation refers to specific gate combinations, such as certain configurations of H and CNOT gates, that can be replaced by a single or fewer gates with equivalent functionality. Gate commutation implies that although the order of quantum gate operations generally affects the final outcome, certain gates—such as CNOT—can be swapped without changing the effect on the quantum state, provided specific conditions are met, as illustrated in Figure \ref{fig:figure1}.

These optimizations exploit the algebraic properties of quantum gates and their representations—such as matrix operations and ZX calculus—to simplify circuit structures, thereby enhancing computational efficiency and reducing error accumulation.

The permutations and combinations of various rule-based transformations can produce a multitude of functionally equivalent circuits, with the total gate count potentially increasing or decreasing. Consequently, the task of reducing the number of two-qubit gates in logical quantum circuits can be formulated as an optimization problem:

\begin{equation}
	\min_{x \in X} F(x)
\end{equation}

Where $F(x)$ denotes the number of two-qubit gates in the circuit, $x$ represents a new equivalent circuit generated via intra-circuit gate transformations using predefined rules, and $X$ denotes the set of all possible resulting circuits. The objective is to find the optimal circuit $x$, minimizing the number of two-qubit gates and thereby achieving the goal of optimizing two-qubit gates.

\subsection{ZX-Calculus}
\subsubsection{ZX diagram rewriting}

ZX calculus is a graphical formalism used to represent linear maps on qubits through ZX diagrams. It uses diagrams to depict operators and quantum states in quantum mechanics, thereby enabling intuitive visual reasoning for complex quantum processes. ZX diagrams are composed of wires, nodes, and a set of transformation rules. Wires represent qubits or the states of quantum systems, and their direction typically corresponds to the flow of time or quantum information. Nodes represent operations on qubits, such as quantum gates. Different node types correspond to specific gates (e.g., H, CNOT), and node shapes or labels are used to distinguish between various operations.

Let $C_{ori}$ denote the initial input quantum circuit. ZX calculus can transform it into a ZX diagram through rewrite rules, i.e.
\begin{equation}
	G_{\mathrm{ori}} = {To\_Graph( C_{\mathrm{ori}}})
\end{equation}

\begin{figure}[h]
	\centering
	\includegraphics[width=60mm,height=30mm]{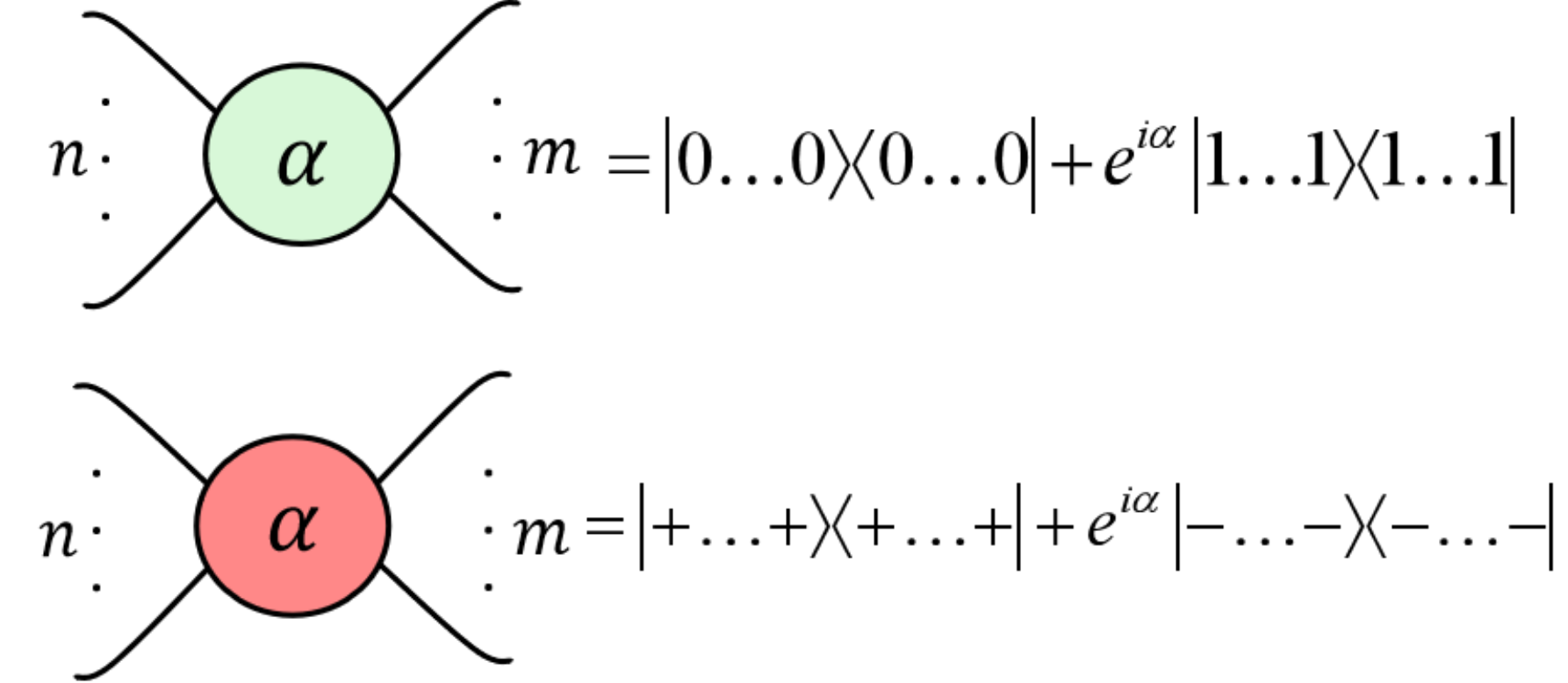}
	\caption{ZX spider nodes}
	\label{fig:figure2}
\end{figure}

The quantum process is represented by a ZX diagram, which consists of wires and spider webs. Spiders can have any number of inputs and outputs, and there are two types: Z spiders (shown in green) and X spiders (shown in red), as illustrated in Figure \ref{fig:figure2}.\par

\begin{figure}[h]
	\centering
	\includegraphics[width=100mm]{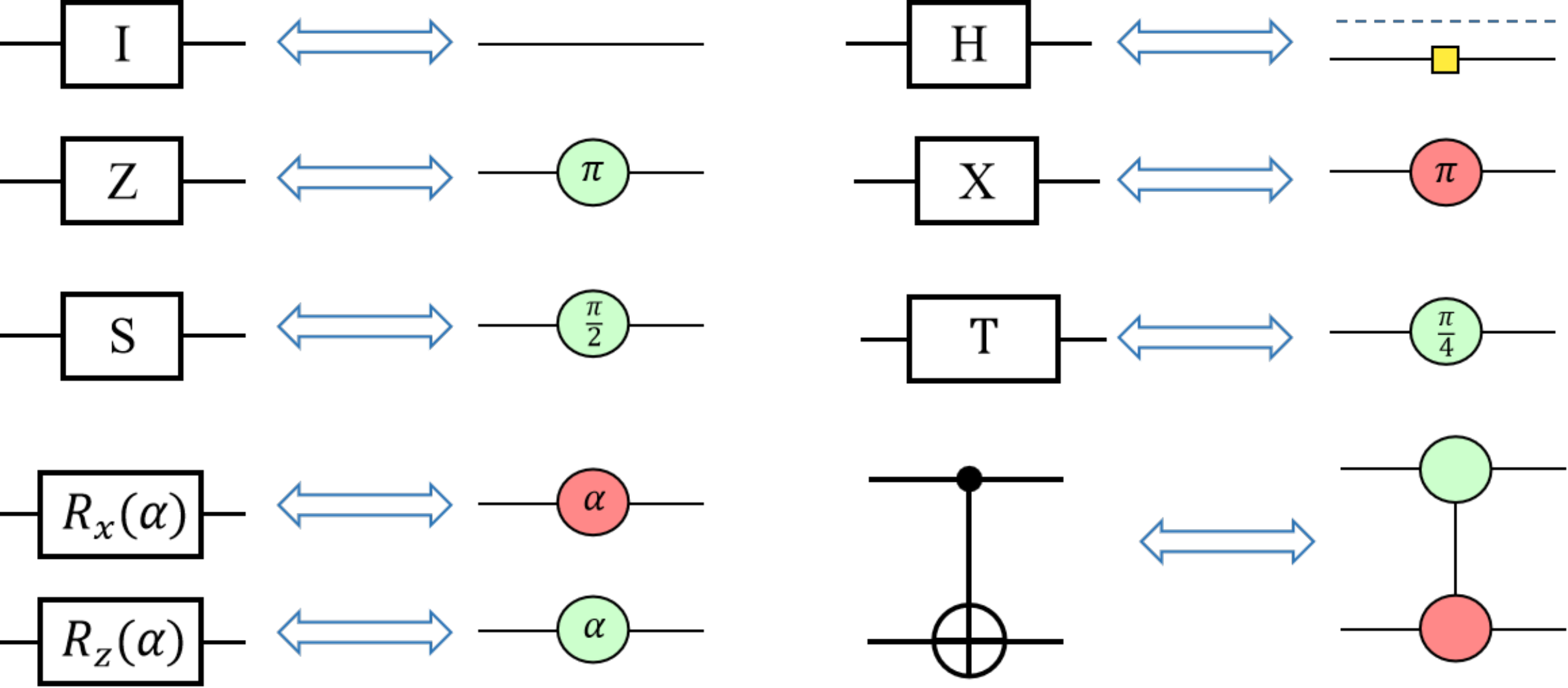}
	\caption{Correspondence between Quantum Gates and ZX Nodes}
	\label{fig:figure3}
\end{figure}

Quantum gates in quantum circuits can be represented as nodes with different colors and shapes and connecting wires in ZX diagrams. The Pauli-Z gate is denoted by a green circle, while the Pauli-X gate is represented by a red circle, with the rotation angle indicated by $\alpha$. The Hadamard (H) gate is depicted by a yellow square or a dashed line. In the case of the CNOT gate, the control qubit is a green node, and the target qubit is a red node, as shown in Figure \ref{fig:figure3}.\par

\begin{figure}[h]
	\centering
	\includegraphics[width=100mm,height=30mm]{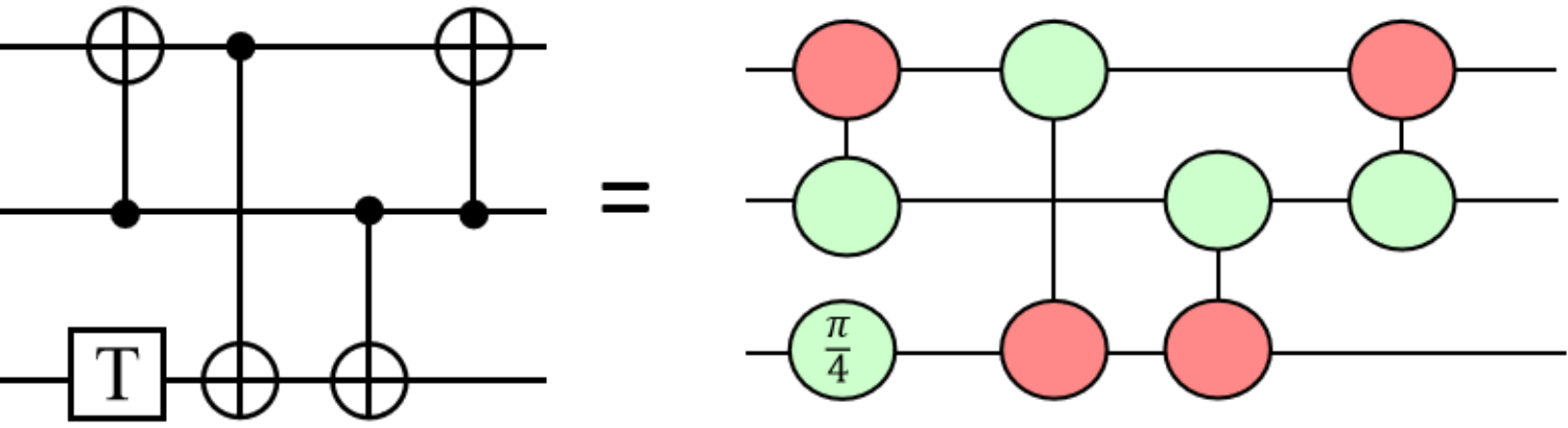}
	\caption{Transform the circuit into a ZX diagram}
	\label{fig:figure4}
\end{figure}

In quantum circuits, any quantum gate can be decomposed into combinations of X and Z spiders in ZX-diagrams according to the node correspondence scheme illustrated in Figure \ref{fig:figure3}. This decomposition enables the transformation of quantum circuits into ZX-calculus graph representations, as demonstrated in Figure \ref{fig:figure4}.\par
\subsubsection{ZX rules}

\begin{figure}[h]
	\centering
	\includegraphics[width=130mm,height=50mm]{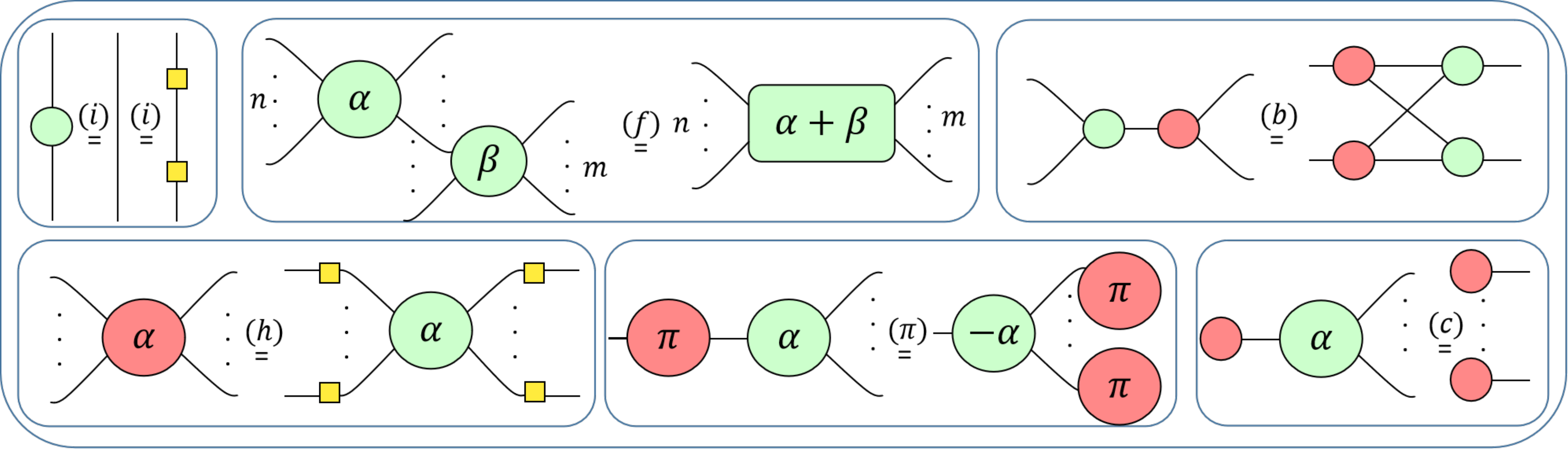}
	\caption{Fundamental Rules of the ZX Calculus}
	\label{fig:figure5}
\end{figure}

ZX calculus defines a set of graphical transformation rules that allow for simplification and reasoning based on fundamental principles of quantum mechanics and quantum computation. In ZX diagrams, only the nodes and their connections are significant; the spatial positions and lengths of the wires are irrelevant and can be stretched or deformed without affecting the diagram’s meaning. Basic transformation rules include node fusion, color change, and wire splitting, which collectively allow a ZX diagram to be simplified into its minimal form containing the fewest possible nodes and edges, as illustrated in Figure \ref{fig:figure5}.\par
\begin{figure}[h]
	\centering
	\includegraphics[width=120mm,height=25mm]{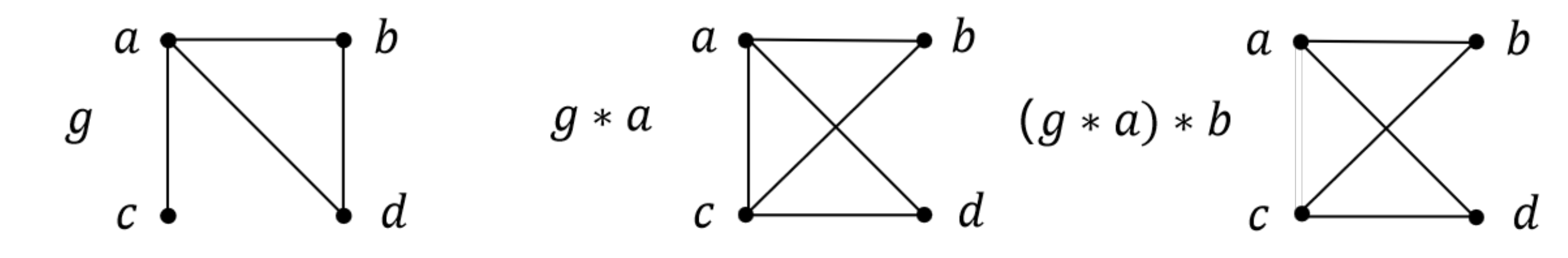}
	\caption{local complementation rule}
	\label{fig:figure6}
	
\end{figure}

In addition to node-level operations, ZX calculus also defines edge-level transformations, notably \textit{local complementation} (lc) and \textit{pivoting}(p). Local complementation simplifies the graphical representation of quantum circuits by locally adjusting spider phases and connectivity. Under certain conditions, it allows for the inversion or transformation of specific parts of the diagram without affecting its global behavior. Given a graph $g$ and a vertex $a \in g$, the local complement of $g$ with respect to $a$, denoted $g * a$, is a graph with the same vertices and edges as $g$, except that the neighborhood of $a$ is complemented. If the phase of spider $a$ in $g * a$ is $\pm \pi/2$, then this phase is subtracted from its adjacent spiders, and spider $a$ is eliminated to simplify the diagram, as illustrated in Figure \ref{fig:figure6}.

\begin{figure}[h]
	\centering
	\includegraphics[width=110mm,height=30mm]{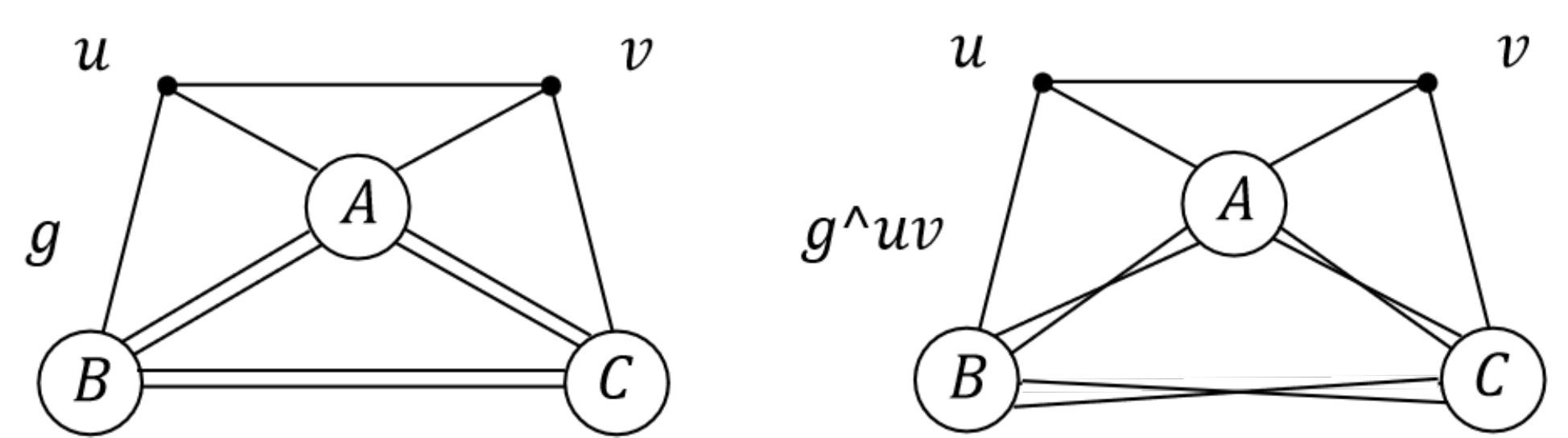}
	\caption{pivot rule}
	\label{fig:figure7}
\end{figure}

The \textit{pivoting rule} can be interpreted as a sequence of three local complementations applied to an edge $(u, v) \in E$. In the resulting pivoted graph $g \land uv$, two vertices from distinct sets $A$, $B$, or $C$ are connected if and only if they were not connected in $g$, while connections within the same set remain unchanged. Comparing $g$ and $g \land uv$, let $A$ be the set of vertices adjacent to both $u$ and $v$, $B$ the set adjacent to $u$ but not to $v$, and $C$ the set adjacent to $v$ but not to $u$. The pivot operation involves swapping $u$ and $v$, and complementing the connections between each pair of sets $A$, $B$, and $C$. A vertex in $A$ will be connected to a vertex in $B$ after the pivot if and only if they were not connected before; the same applies to the other two set pairs. All remaining edges, including those within sets $A$, $B$, and $C$, remain unchanged, as shown in Figure \ref{fig:figure7}.

\begin{figure}[h]
	\centering
	\includegraphics[height=40mm]{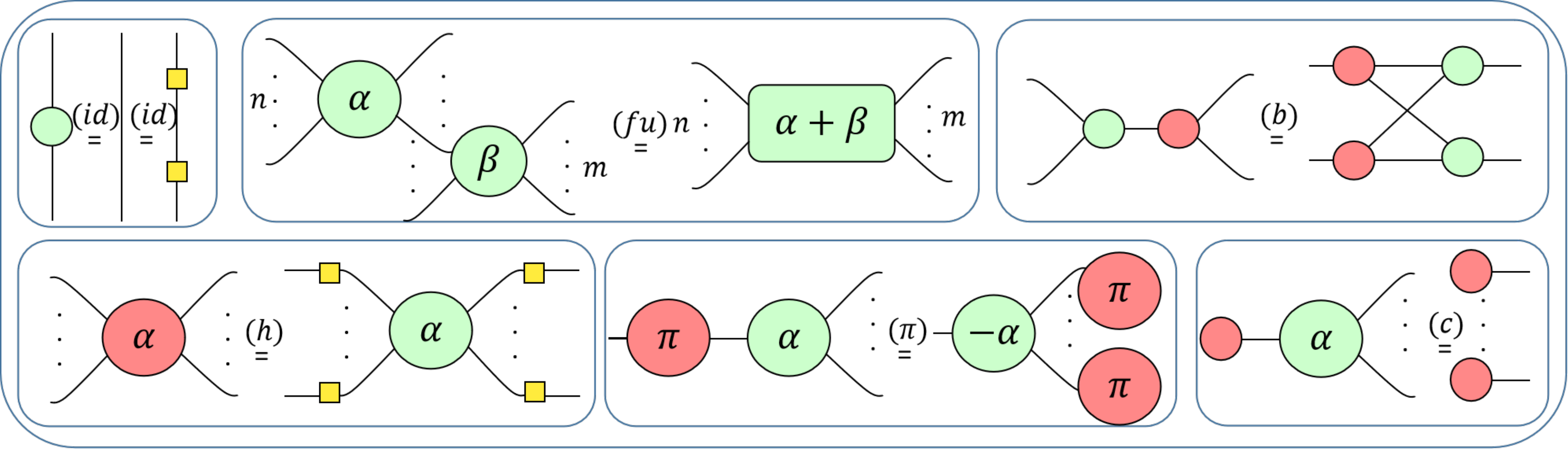}
	\caption{An Example of Optimization Using ZX Diagrams}
	\label{fig:figure8}
\end{figure}

A ZX diagram can be simplified into a more compact structure through a sequence of ZX calculus transformations. First, the fusion rule ($fu$) is applied multiple times to merge adjacent nodes of the same type, thereby reducing the overall node count. Next, the Hadamard color change rule ($h$) is employed to transform all red (X-type) spiders into green (Z-type) spiders, unifying the diagram’s node types. Subsequently, pivoting transformations are performed between nodes of the same color to restructure the connectivity and minimize the number of Hadamard edges. Finally, an additional round of node fusion is conducted, and some green spiders are color-inverted to further reduce the number of Hadamard gates introduced during the diagram-to-circuit extraction process. These steps collectively enable the derivation of an optimized quantum circuit with a simplified structure and fewer quantum gates, as shown in Figure \ref{fig:figure8}.
\subsubsection{Circuit extraction}

The ZX diagram obtained after simplification is a compact graphical representation that cannot be directly used as a quantum circuit. Therefore, a circuit extraction procedure is required to convert the simplified ZX diagram into an executable quantum circuit. Backens \textit{et al.}~\cite{backensThereBackAgain2021a} proposed an extraction algorithm capable of transforming ZX diagrams that possess the \textit{generalized flow} (gflow) property into quantum circuits. Gflow is a combined structural property of ZX diagrams and the measurement-based quantum computation (MBQC) model, which defines a rule for assigning a measurement order and measurement angle to each qubit in an open graph.

\begin{figure}[h]
	\centering
	\includegraphics[width=110mm,height=20mm]{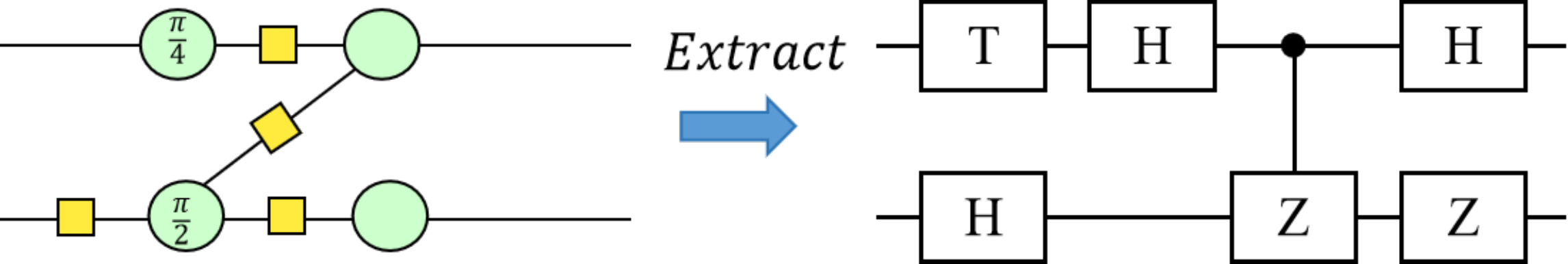}
	\caption{An Example of Extracting ZX Diagrams}
	\label{fig:figure9}
\end{figure}

Let $G_{\mathrm{ori}}$ denote the optimized ZX diagram with gflow properties, which can be extracted into circuit $C_{\mathrm{extract}}$ as:
\begin{equation}
	C_{\mathrm{extract}}=Extract(G_{\mathrm{ori}})
\end{equation}
The extraction of a quantum circuit from a ZX diagram follows a right-to-left traversal strategy, where vertices are progressively consumed from the output side of the graph and translated into quantum gates placed toward the input side. At each step, vertices in the maximum-delay layer are identified and eliminated through local rewrite operations. This process incrementally reconstructs an equivalent quantum circuit by introducing standard gate primitives such as CNOT, Hadamard, and phase gates. The goal is to transform the ZX diagram into a gate-level circuit while preserving its semantic equivalence, as illustrated in Figure \ref{fig:figure9}.

\section{Quantum Circuit Optimization Scheme Based on Dynamic Grouping and ZX Calculus}

\subsection{Problem Definition and Scheme Overview}

Given a quantum circuit $QC(G,Q)$, where $G=\left\{g_1,g_2,{\cdots,g}_{N_G}\right\}$ represents the set of $N_G$ quantum gates and $Q = \{q_1, q_2, \cdots, q_{N_Q}\}$ denotes the set of $N_Q$ qubits, the objective is to develop a strategy for partitioning the circuit $QC$ into $N_S$ sub-circuits ${QC}_i^s$ ($i \in \{1, 2, \cdots, N_S\}$). These sub-circuits are subsequently subjected to equivalence screening using ZX calculus to derive optimized sub-circuits ${QC}_i^{s^\prime}$ ($i \in \{1, 2, \cdots, N_S\}$), which are synthesized into a final circuit $QC'$ with a minimized number of two-qubit gates.\par

\begin{figure}[h]
	\centering
	\includegraphics[width=150mm]{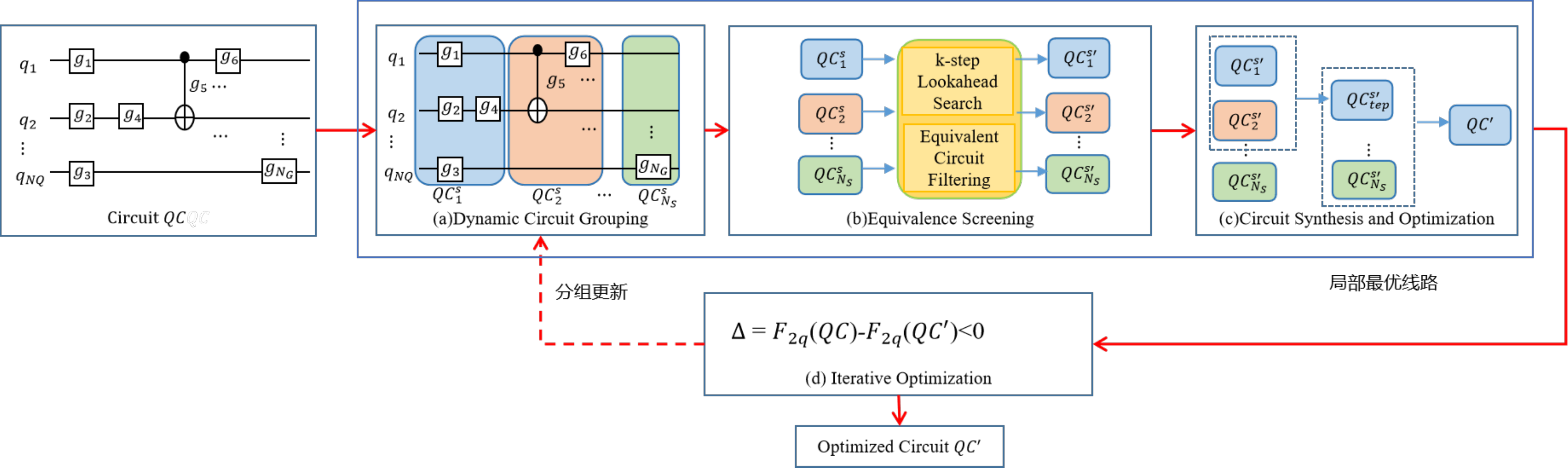}
	\caption{Framework Diagram of Quantum Circuit Optimization via Dynamic Grouping and ZX Calculus}
	\label{fig:figure3-1}
\end{figure}

To address this problem, we propose a quantum circuit optimization scheme based on dynamic grouping and ZX calculus. The scheme utilizes simulated annealing for multi-round iterative optimization, where each round consists of three steps: dynamic grouping of the circuit based on randomized strategies, sub-circuit equivalence screening using ZX calculus combined with a $k$-step lookahead search, and circuit synthesis optimization via delayed placement, as illustrated in Figure~\ref{fig:figure3-1}.

\begin{enumerate}[(1)]
	
	\item \textbf{Circuit Dynamic Grouping Method Based on Stochastic Strategy:} \\
	For each gate $g_i$ in the circuit $QC(Q, G)$, the gate depth $l_{g_i}$ is first computed. Gates with the same depth are grouped into a set $L_t$, where $L_t = \{g_i \mid l_{g_i} = t, t \in \{1, 2, \cdots, d\}\}$. This process yields a $d$-layer gate collection $L = \{L_1, L_2, \cdots, L_d\}$. Next, the number of groups $N_S$ is randomly determined, and the length of each group $m_i$ is computed based on the total number of layers. This forms a grouping structure $M_S = \{m_1, m_2, \cdots, m_{N_S}\}$. Finally, the gate layers in $L$ are grouped according to $M_S$, where consecutive layers of length $m_i$ are merged to form sub-circuits ${QC}_i^s$.
	
	\item \textbf{Subcircuit Equivalence Selection Method Based on ZX Calculus and Look-Ahead Search:} \\
	For each sub-circuit ${QC}_i^s$, ZX calculus is first used to transform the circuit into a ZX diagram $D_i^S$. Node fusion and color-change rules are then applied to obtain a preprocessed diagram $D_0$. A $k$-step lookahead search is performed to identify optimal transformations over future steps. This process involves four stages: 
	Generate a set of matching candidate rules $R$ for $D_0$; Randomly select $j$ rules from $R$ and apply them to produce a set of candidate transformed diagrams $I^1 = \{I_1^1, I_2^1, \cdots, I_j^1\}$, and select the diagram $D_1$ with the fewest Hadamard wires; Apply $k$ rounds of rule matching and transformation on $D_1$ to generate a candidate set of diagrams $D=\{D_1, D_2, \cdots, D_k\}$, from which the one with the fewest two-qubit gates is selected as $D_k^\prime$; Repeat the $k$-step lookahead process until no further beneficial transformations can be applied or all rules are exhausted, resulting in the optimized diagram $D_i^{S_{Kstep}^\prime}$. Finally, $D_i^{S_{Kstep}^\prime}$ is extracted back into an equivalent sub-circuit ${QC}_i^{S^\prime}$. After applying this process to all sub-circuits, the optimized set of sub-circuits ${QC}^{S^\prime}$ is obtained.

	\item \textbf{Circuit Synthesis and Optimization Method Based on Delayed Placement:} \\
	Given the optimized sub-circuit set ${QC}^{S^\prime}$, each sub-circuit ${QC}i^{S^\prime}$ is iteratively merged to construct a new circuit ${QC}_{\text{merge}}$. During this process, the ${QC}_{\text{merge}}$ circuit is simplified using the $F_{BO}$ method. Delayed gate placement is further employed to expose more gate cancellation opportunities, thereby producing the optimized circuit $QC^\prime$.
	
	\item \textbf{Iterative Circuit Optimization Method Based on Simulated Annealing:} \\
	Given a quantum circuit $QC$, the proposed methods (1)–(3) produce an optimized circuit $QC^\prime$. The difference in the number of two-qubit gates between $QC$ and $QC^\prime$ is defined as $\Delta$, which is treated as the energy change in the simulated annealing framework. Based on the Metropolis criterion, $QC^\prime$ is accepted as the new state if it satisfies the acceptance condition. The optimization process is iterated, with $QC^\prime$ used as the new starting circuit, until the system reaches the temperature threshold or no further improvements can be found. This yields a globally optimized solution.
	
\end{enumerate}
\subsection{Dynamic Circuit Grouping Method Based on Random Strategy}
\begin{figure}[h]
	\centering
	\includegraphics[height=40mm]{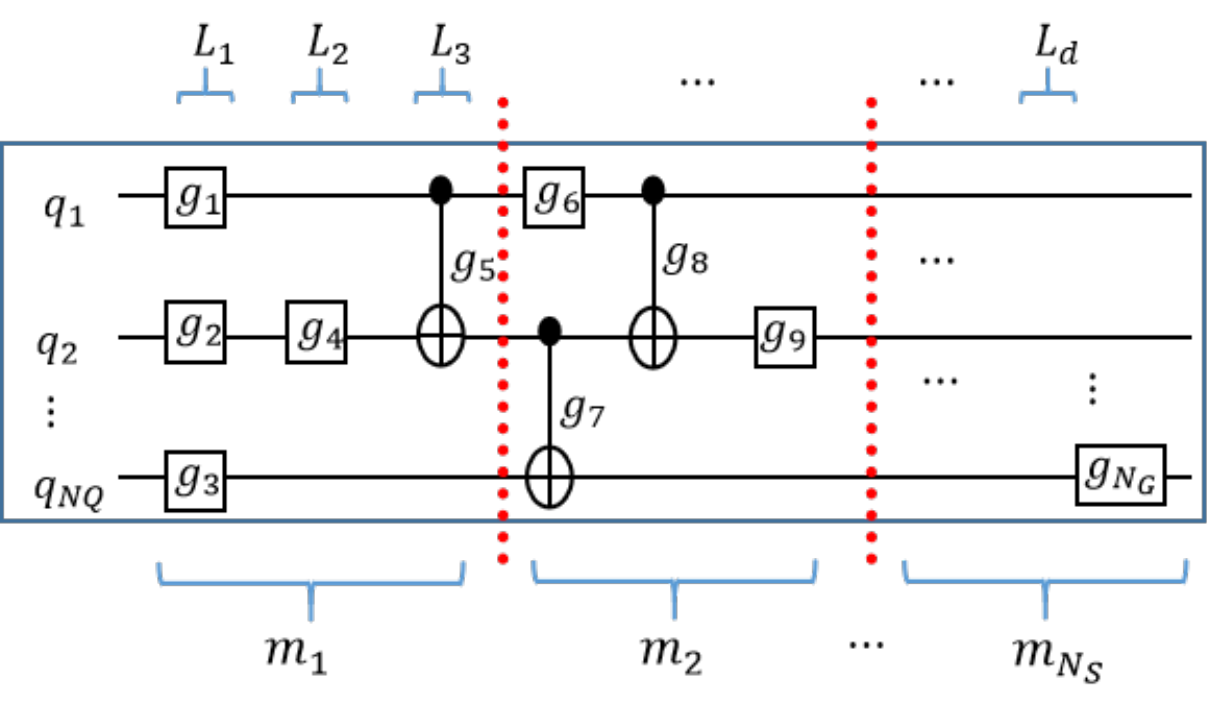}
	\caption{Dynamic Circuit Grouping}
	\label{fig:figure3-2}
\end{figure}

To achieve grouping of circuit $QC$, a dynamic circuit grouping method based on random strategies is designed, comprising three components: (1) circuit layer partitioning, (2) grouping parameter calculation, and (3) sub-circuit division, as illustrated in Figure \ref{fig:figure3-2}.
\subsubsection{Circuit Layer Partitioning}
For each gate $g_i$ in the quantum circuit $QC(Q, G)$, the layer index $l_{g_i}$ must be computed. Gates with the same layer index are grouped into the same layer set $L$. To assign layers, the depth of each quantum gate is determined by considering all other gates acting on the same qubit as $g_i$, since quantum gates are executed sequentially.

Let $G_q(g_i)$ denote the set of all preceding gates acting on the same qubit as $g_i$:
\begin{equation}
	G_q({g_i})=\{\left.g_k\right|q_{g_k}=q_{g_i},k\in(1,2,\cdots,i)\}
\end{equation}

Then, the layer index $l_{g_i}$ is calculated as follows:

\begin{equation}
	l_{g_i} = \begin{cases} 
		\max\, l(G_q(g_i)) + 1, & G_q(g_i) \neq \varnothing \\ 
		1, & G_q(g_i) = \varnothing 
	\end{cases}
\end{equation}
The layer $l_{g_i}$ of gate $g_i$ in the circuit is computed based on whether $G_q({g_i})$ (the set of gates on the same qubit as $g_i$) contains prior gates. If $G_q(g_i)$ is empty, then $g_i$ is placed in the first layer. Otherwise, its layer index is assigned as one greater than the maximum depth of all gates in $G_q(g_i)$. For a two-qubit gate $g_i$, the maximum of the layer indices for its control and target qubits is used. For two-qubit gates, the layer indices of both the control and target qubits are computed, and the larger of the two is assigned as the gate's layer. Finally, all gates are grouped according to their layer indices. This iterative process results in a total of $d$ layers, reorganizing all gates $g_i \in \{1, 2, \cdots, N_G\}$ into the layered structure $L = \{L_1, L_2, \cdots, L_d\}$.

\subsubsection{Grouping Parameter Calculation}
To determine grouping parameters, a random method is used to compute the number of groups $N_S$, followed by calculating the length of each group $M_S$. The grouping process employs a random pattern, with the minimum number of groups set to a constant b and the maximum number of groups set to $d/N_Q$, where $d$ is the total number of layers and $N_Q$ is the number of qubits. The number of groups $N_S$ is generated via a random grouping strategy, expressed as:

\begin{equation}
	N_S=random(b,\frac{d}{N_Q})\ \ 
\end{equation}

After determining the number of groups $N_S$, the length of each group mi is randomly generated, with the constraint that the sum of all group lengths equals the total number of layers $d$.
\begin{equation}
	m_i = \text{random} \left( a, \frac{d}{n^{(r)} - 1} \right), \quad \text{s.t.} \sum_{i=1}^{N_s} m_i = d
\end{equation}

At this stage, the partitioning configuration $M_S = \{m_1, m_2, \cdots, m_{N_S}\}$ is obtained through a randomized partitioning strategy.
\subsubsection{Circuit Partitioning}

The circuit $QC$ is represented as a layered gate set $L = \{L_1, L_2, \cdots, L_d\}$. Based on the grouping configuration $M_S = \{m_1, m_2, \cdots, m_{N_S}\}$, the $d$ layers in $L$ are partitioned into $N_S$ groups, where each group contains $m_i$ consecutive layers. These layer combinations form sub-circuits ${QC}_i^s$.

\begin{align}
	QC_i^s &= \left\{ L_{1+\sum_{1}^{i-1} m_i}, L_{2+\sum_{1}^{i-1} m_i}, \cdots, L_{\sum_{1}^{i} m_i} \right\}, \quad i \in \{1, 2, \cdots, N_S\}
\end{align}

The term $\sum_{1}^{i-1} m_j$ represents the total number of layers in the first $i-1$ groups, and the sub-circuit ${QC}_i^s$ contains all gates from layer $\sum_{1}^{i-1} m_j + 1$ to layer $\sum_{1}^{i} m_j$.\par
\begin{figure}[h]
	\centering
	\includegraphics[width=120mm]{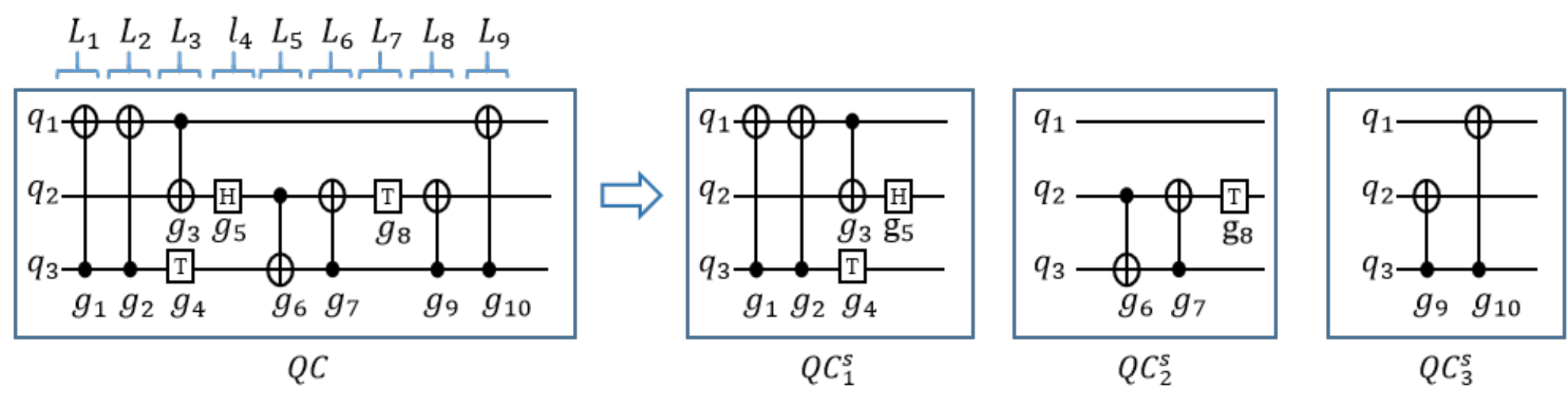}
	\caption{Example of Circuit Grouping}
	\label{fig:figure3-3}
\end{figure}

As an illustrative example, consider a circuit with 3 qubits and 10 gates. After performing layer assignment, the gates are distributed across 9 layers: $L_1 = {g_1}$, $L_2 = {g_2}$, $L_3 = \{g_3, g_4\}$, $L_4 = {g_5}$, $L_5 = {g_6}$, $L_6 = {g_7}$, $L_7 = {g_8}$, $L_8 = {g_9}$, and $L_9 = {g_{10}}$. Using a randomized grouping strategy, the circuit is divided into 3 groups with respective lengths of 4, 3, and 2 layers. The circuit is then segmented into corresponding sub-circuits as illustrated in Figure \ref{fig:figure3-3}.

\subsection{Sub-Circuit Equivalence Screening Method Based on ZX Calculus and $k$-Step Lookahead Search}

\begin{figure}[h]
	\centering
	\includegraphics[width=120mm]{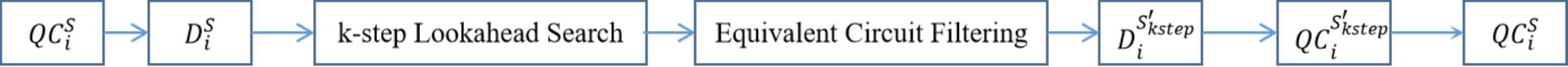}
	\caption{Subcircuit Optimization and Selection Flow}
	\label{fig:figure3-4}
\end{figure}

Following dynamic grouping, circuit $QC$ is partitioned into $N\_S$ sub-circuits $QC_i^s$ $(i \in {1,2,…,N_S})$. To optimize these sub-circuits for equivalence, a method combining ZX calculus and $k$-step lookahead search is proposed. The method consists of three steps: (1) diagram conversion, (2) $k$-step lookahead equivalence screening, and (3) circuit extraction. The overall workflow is illustrated in Figure \ref{fig:figure3-4}.
\subsubsection{Diagram conversion}

The sub-circuit ${QC}_i^S$ is converted into a ZX diagram denoted as $D_i^S = (V, E)$, where $V$ represents the set of nodes, and $i$ indexes the sub-circuit. The set $E$ describes the edges in the diagram; each edge connects two nodes $v$ and $u$, and is annotated with a label $et$ indicating the edge type.
\begin{equation}
	D_i^S=To\_Graph(QC_i^S)
\end{equation}
Node fusion rules are applied to $D_i^S$ to simplify the structure by merging nodes. Subsequently, the $To\_Graph$ method is used to transform all X-spiders into Z-spiders, resulting in the preprocessed diagram $D_0$.

\subsubsection{$K$-Step Lookahead Transformation}

\begin{figure}[h]
	\centering
	\includegraphics[width=120mm]{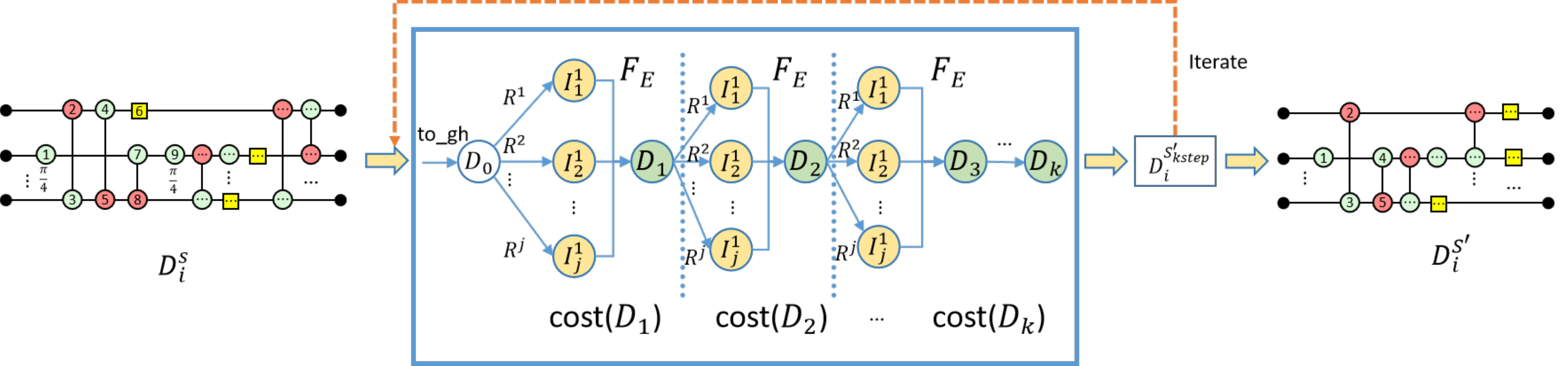}
	\caption{$k$-Step Look-Ahead Equivalence Selection}
	\label{fig:figure3-5}
\end{figure}

Equivalence transformations using ZX rules are applied to $D_0$, with the $k$-step lookahead screening algorithm employed to identify optimal transformations. The goal is to find the equivalent diagram minimizing the number of two-qubit gates in the extracted circuit. The algorithm consists of four components: 1) candidate rule set generation, 2) single-step ZX diagram equivalence screening, 3) $k$-step lookahead screening, and 4) iterative search optimization, as shown in Figure \ref{fig:figure3-5}.

Given a quantum circuit $QC_i^S$, the diagram $D_0$ is obtained through diagram conversion and rule preprocessing. A 3-step lookahead equivalence screening is performed on $D_0$:

\begin{enumerate}[(1)]
	\item ZX diagrams allow rule matching via a matching function $R(D)$ that identifies applicable transformation rules within a circuit. Specifically, $R_{id}$, $R_{fu}$, $R_{lc}$, and $R_p$ denote the matching results corresponding to identity, fusion, local complementation, and phase rules, respectively. Each matched rule can be applied using the rule transformation function $F_R$, which simplifies the ZX diagram by reducing node and edge complexity. The functions $\mathrm{LCH}(G)$ \cite{staudacherReducing2QuBitGate2023a}and $\mathrm{PH}(G)$ \cite{staudacherReducing2QuBitGate2023a} are used to identify matchable nodes for local complementation and phase rules in diagram $G$. Applying these to the preprocessed diagram $D_0$ yields rule match sets $R_{LCH}$ and $R_{PH}$, respectively. Their union forms the candidate transformation set $R = \{R_1, R_2, \cdots, R_{N_R}\}$ of size $N_R$. Rules are applied using the function $\mathrm{apply}(D, R(D))$.
	
	\item From the candidate set $R$, a rule $R_I$ is randomly selected for application to $D_0$:
	\begin{equation}
		R_I = R_i, \quad \text{where} \quad i \sim \mathcal{U}(1, N_R)
	\end{equation}
	This results in a candidate diagram $I_1^1 = F_R(D_0, R_I)$. To enhance simplification, additional applications of the identity ($R_{id}$) and fusion ($R_{fu}$) rules are applied to $I_1^1$, further reducing its node count. This random sampling process is repeated $j$ times, resulting in a set of single-step transformed diagrams $I^1 = \{I_1^1, I_2^1, \cdots, I_j^1\}$. The number of Hadamard edges, which impact the final number of quantum gates, is a key metric in evaluating these diagrams. Given a diagram $D_{QC} = (V, E)$, the number of Hadamard edges is computed as:
	\begin{equation}
		F_E(I_j^1) = 
		\begin{cases} 
			e+1 \in E_{e+1}, & \text{ if } et(v,u) \text{ type is Hadamard} \\
			e,  & \text{otherwise}
		\end{cases}
	\end{equation}
	The candidate set $I^1$ is filtered by selecting the diagram with the fewest Hadamard edges:
	
	\begin{equation}
		D_1={\arg\min}_{I_i^1\in I^1}\ F_E(I_i^1)\ \ 
	\end{equation}
	
	\item Starting from $D_1$, the above two steps are recursively applied for $k$ iterations, yielding a sequence of transformed diagrams $\{D_1, D_2, \cdots, D_k\}$. While these transformations often reduce the number of nodes and edges in the ZX diagram, such structural simplifications may not always lead to circuits with fewer two-qubit gates after extraction. Hence, an additional evaluation based on the number of two-qubit gates is introduced.\\
	
	The cost of a circuit $C$ is defined as the number of two-qubit gates:
	\begin{equation}
		\text{cost}(C) = 
		\begin{cases} 
			1, & \text{if } g \in G \text{ and } g \text{ act on 2 qubits} \\
			0, & \text{otherwise}
		\end{cases}
	\end{equation}
	\item The $k$-step lookahead process is repeated until one of the following termination conditions is met: (i) the candidate transformation set $R$ becomes empty, or (ii) no improvement is observed in the cost metric for $p = 5$ consecutive iterations. The final diagram $D_k^\prime$ is then considered the optimized result.

\end{enumerate}

\subsubsection{Circuit Extract}
The optimized diagram $D_k^\prime$ is converted back into the sub-circuit ${QC}_i^{S^\prime}$, which minimizes the number of two-qubit gates.

\begin{equation}
	{QC}_i^{S^\prime}=Extract(D_k^\prime)\ 
\end{equation}

\begin{figure}[h]
	\centering
	\includegraphics[width=150mm]{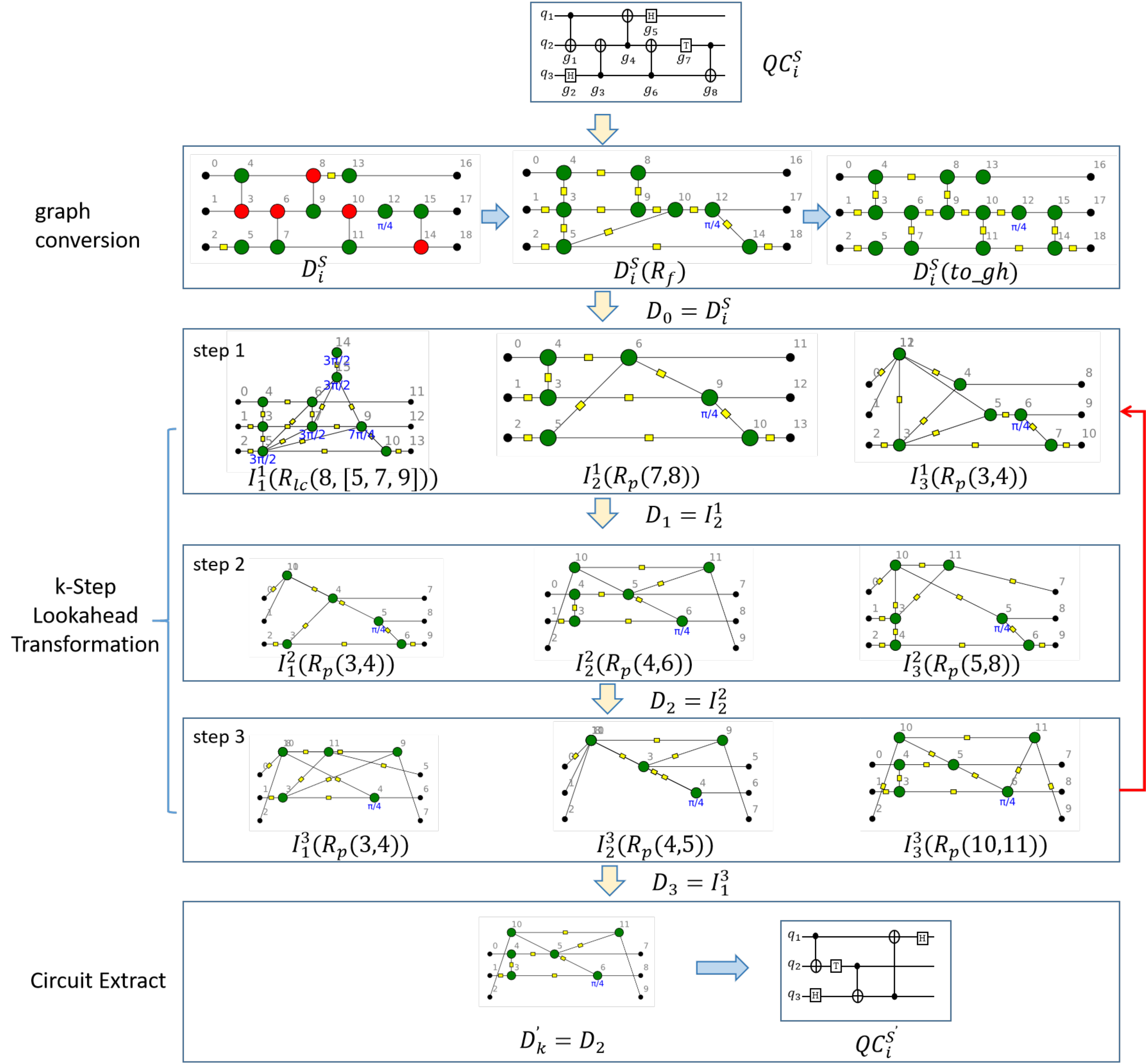}
	\caption{Example of Equivalence Selection Based on $k$-Step Lookahead}
	\label{fig:figure3-6}
\end{figure}

Consider a sub-circuit ${QC}_i^S$, which is transformed into a ZX diagram $D_0$ via rule preprocessing. A three-step lookahead equivalence optimization is then applied. In the first step, the rules $R{lc}(8,[5,7,9])$, $R_p(7,8)$, and $R_p(3,4)$ are applied to $D_0$, generating three candidate diagrams: $I_1^1$, $I_2^1$, and $I_3^1$, which contain 17, 10, and 11 Hadamard edges, respectively. The diagram with the fewest Hadamard edges, $I_2^1$, is selected as the new diagram $D_1$. In the second step, rules $R_p(3,4)$, $R_p(4,6)$, and $R_p(5,8)$ are applied to $D_1$, producing candidates $I_1^2$, $I_2^2$, and $I_3^2$, with 9, 7, and 12 Hadamard edges, respectively. Again, the optimal diagram $I_2^2$ is selected as $D_2$. The third step involves the application of $R_p(3,4)$, $R_p(4,5)$, and $R_p(10,11)$ to $D_2$, generating $I_1^3$, $I_2^3$, and $I_3^3$, with corresponding Hadamard edge counts of 8, 10, and 10. The best result, $I_1^3$, is chosen as $D_3$. The two-qubit gate counts of the extracted circuits from $D_1$, $D_2$, and $D_3$ are then evaluated, yielding 4, 3, and 4 gates, respectively. Based on this metric, $D_2$ is identified as the optimal diagram within the three-step search and is set as the updated $D_0$ for continued rule matching. This process is iteratively repeated for up to $p = 5$ rounds or until no further improvement is observed. The final optimized diagram $D_k^\prime$ is extracted to generate the corresponding sub-circuit ${QC}_i^{S^\prime}$, as shown in Figure \ref{fig:figure3-6}.

\subsection{Circuit Synthesis Optimization Method Based on Delayed Placement}
\begin{figure}[h]
	\centering
	\includegraphics[width=100mm]{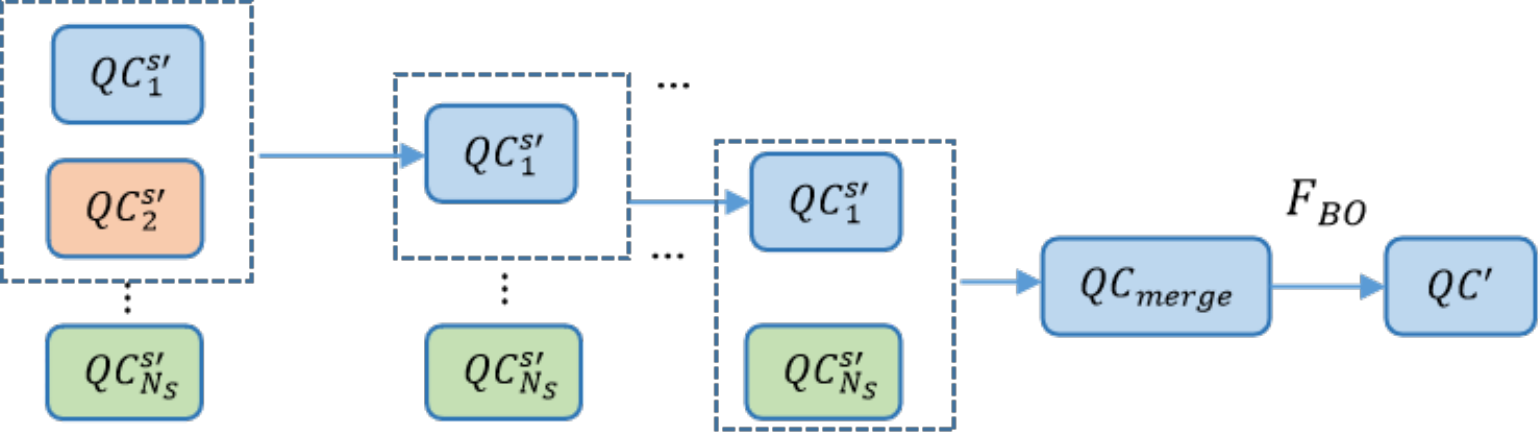}
	\caption{Circuit Synthesis}
	\label{fig:figure3-7}
\end{figure}

After applying lookahead-based equivalence screening, each sub-circuit ${QC}_i^S$ is transformed into an optimized sub-circuit ${QC}_i^{S^\prime}$ that minimizes the number of two-qubit gates. These optimized sub-circuits are then combined to form a global circuit ${QC}^{S^\prime}$. However, due to the independent optimization of each group, there may still exist gate-level redundancies between adjacent sub-circuits. To address this, we propose a post-processing optimization method based on delayed gate placement. This procedure includes two main stages: (1) circuit synthesis and (2) delayed placement optimization, as illustrated in Figure \ref{fig:figure3-7}.\par
Given two sub-circuits $c_1$ and $c_2$, where $G^{c_1} = \{g_1^{c_1}, \cdots, g_{N_{c_1}}^{c_1}\}$ and $G^{c_2} = \{g_1^{c_2},  \cdots, g_{N_{c_2}}^{c_2}\}$ represent their respective sets of quantum gates, the merged circuit $C_{\text{merge}}$ is constructed as:
\begin{equation}
	C_{merge}(c_1,c_2)=\{g_1^{c_1},g_2^{c_1},\cdots,g_{N_{c_1}}^{c_1},g_1^{c_2},g_2^{c_1},\cdots,g_{N_{c_2}}^{c_2}\}\ \ 
\end{equation}

For the complete optimized set ${QC}^{S^\prime} = \{{QC}_1^{S^\prime}, {QC}_2^{S^\prime}, \cdots, {QC}_{N_S}^{S^\prime}\}$, gates are merged sequentially by aligning gates acting on the same qubits into the same rows. This iterative process results in a temporary composite circuit ${QC}_{{merge}} = C_{{merge}}({QC}_1^{S^\prime}, {QC}_2^{S^\prime}, \cdots, {QC}_{N_S}^{S^\prime})$ containing $N_M$ gates.

In the circuit extraction process from ZX diagrams, additional Hadamard gates may be introduced to preserve equivalence, leading to suboptimal gate configurations across merged segments. Therefore, a further optimization step is necessary. We adopt the basic\_optimization strategy $F_{BO}$ proposed in \cite{kissingerPyZXLargeScale2020}, which applies rule-based transformations to reduce circuit depth and gate count by delaying gate placements to uncover more cancellation opportunities. The final optimized circuit is denoted as:
\begin{equation}
	{QC}^\prime=F_{BO}({QC}_{merge}^\prime) 
\end{equation}
This optimization involves rule matching based on known gate commutation and combination patterns. For each qubit $q \in Q$, a gate stack $S_q$ is maintained. When a gate $g_i$ acting on $q$ is encountered, it is pushed onto $S_q$. If $g_i$ is successfully merged with a previous gate in the stack, it is removed. During traversal, a decision is made to place $g_i$ immediately or delay its placement. If $g_i$ can be canceled with a gate $g_s \in S_q$, the cancellation is performed immediately. If $g_i$ commutes with but cannot yet be canceled with gates in $S_q$, it is retained in the stack to await future opportunities. The entire circuit is iteratively traversed in both forward ($g_1 \rightarrow g_{N_M}$) and backward ($g_{N_M} \rightarrow g_1$) directions until no further reduction in gate count is observed.

\begin{figure}[h]
	\centering
	\includegraphics[width=150mm]{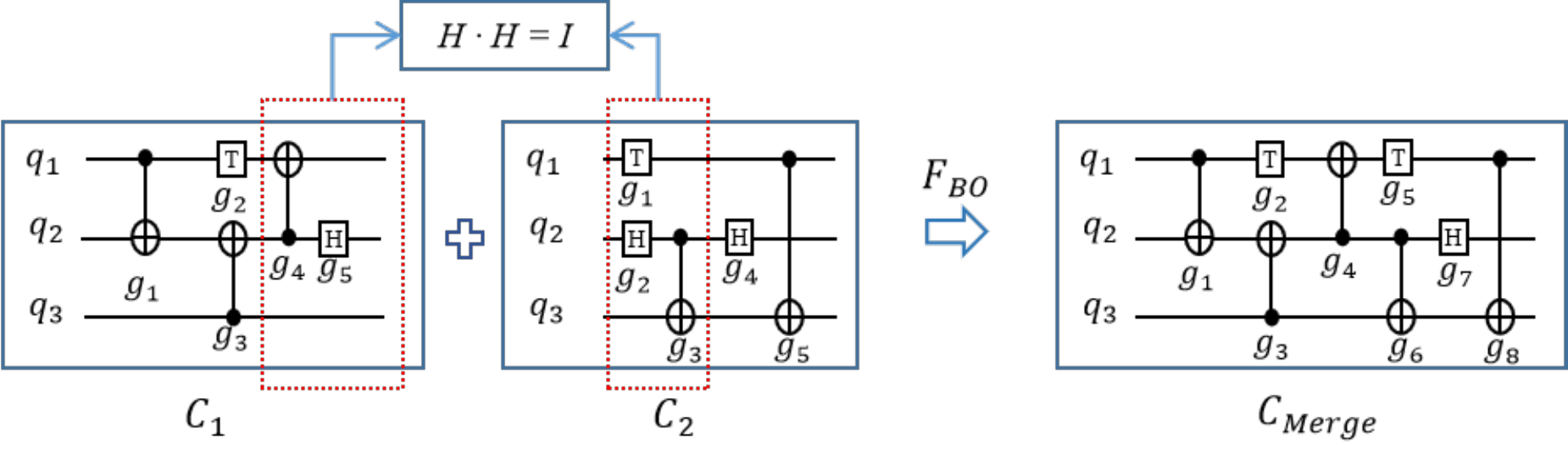}
	\caption{Example of Circuit Synthesis Optimization}
	\label{fig:figure3-8}
\end{figure}

As an example, consider two 3-qubit circuits $C_1$ and $C_2$, each containing five gates. The gates of $C_2$ are appended to $C_1$ to form a temporary circuit $C_{\text{tep}}$. Applying $F_{BO}$ to $C_{\text{tep}}$ identifies that gate $g_5$ in $C_1$ and gate $g_2$ in $C_2$ are consecutive Hadamard gates on the same qubit and can be canceled. The resulting optimized circuit $C_{\text{Merge}}$ is shown in Figure \ref{fig:figure3-8}.

\subsection{Circuit Iterative Optimization Method Based on Simulated Annealing}
\begin{figure}[h]
	\centering
	\includegraphics[width=150mm]{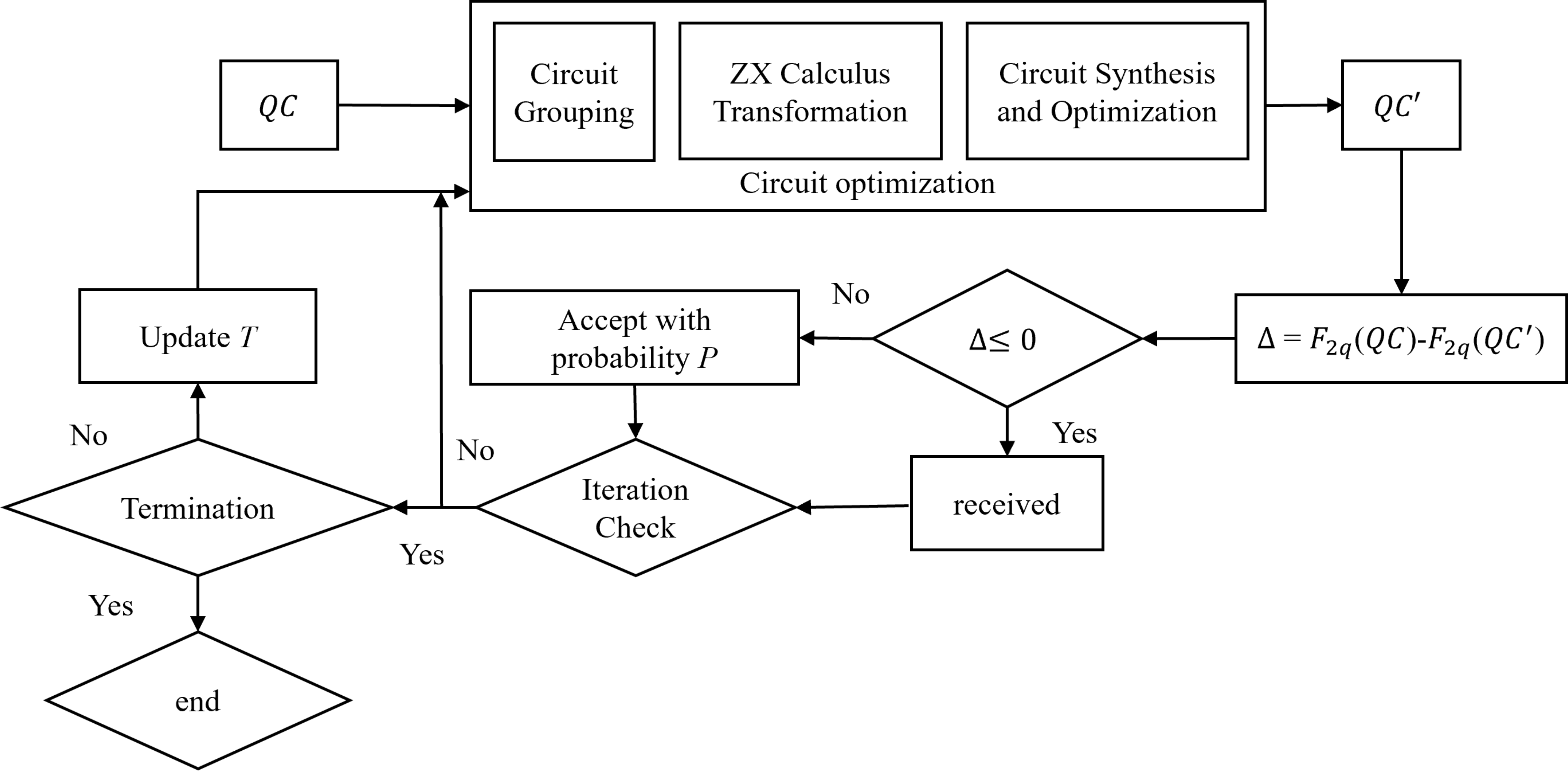}
	\caption{Iterative optimization flow for quantum circuits}
	\label{fig:figure3-9}
\end{figure}

The sub-circuits ${QC}_i^S$ are merged and further optimized through delayed placement to obtain the final circuit ${QC}^\prime$ with a reduced number of two-qubit gates. However, since different grouping strategies may yield varying optimization outcomes, a global coordination mechanism is required to approximate the optimal solution. To this end, we employ a simulated annealing algorithm to iteratively update the grouping configuration and optimize the circuit structure. The annealing process is governed by a temperature parameter $T$, and begins with an initial random grouping. As shown in Figure \ref{fig:figure3-9}:\par
At each iteration, a new grouping configuration is randomly selected, and the best result obtained is retained as a candidate for the next iteration. The detailed procedure is as follows:
\begin{enumerate}[(1)]
	\item Initial Random Grouping:
	Given an arbitrary quantum circuit $QC$, a random grouping strategy is applied to decompose it into a set of sub-circuits ${QC}^S=\{{QC}_1^s,{\cdots,QC}_{N_S}^s\}$. This partitioning is performed randomly during each iteration, generating a candidate grouping ${{QC}_i^s}$ where $i \in \{1, \cdots, N_S\}$.
	
	\item ZX Calculus Transformation:
	Each sub-circuit ${QC}i^s$ is then converted into a ZX diagram, and equivalence optimization is performed using a $k$-step lookahead strategy to generate the optimized sub-circuit ${QC}_{i_{best}}^{s\prime}$.
	
	\item Circuit Synthesis and Optimization:
	All optimized sub-circuits are collected into a set ${QC}^{S^\prime}$, which is synthesized into a single global circuit. This merged circuit is further optimized using the delayed placement strategy to obtain the candidate circuit ${QC}^\prime$.
	
	\item Two-Qubit Gate Evaluation:
	The optimization result is evaluated by comparing the number of two-qubit gates before and after the transformation. Specifically, the difference is computed as $\Delta = F_{2q}(QC) - F_{2q}(QC^\prime)$. If $\Delta \leq 0$, the new circuit represents an improvement and is accepted as the current best solution, with $QC$ updated to $QC^\prime$. If $\Delta > 0$, the new solution is worse, but may still be accepted with a probability $P$, in accordance with the Metropolis criterion.
	
	\item Iteration Check:
	The algorithm checks whether the maximum number of iterations $T$ has been reached. If not, a new grouping is generated and the process is repeated.
	
	\item Termination Condition: 
	A termination condition is defined such that if no improved solution is found after five consecutive iterations following the discovery of a locally optimal solution, the current best circuit $QC^\prime$ is accepted as the global optimum and the optimization process terminates. Otherwise, the temperature is decreased and the iteration count is reset.
	
	\item  Iterative Optimization:
	Steps (1) through (6) are repeated until the termination condition is satisfied, yielding the final optimized circuit $QC^\prime$.
	
\end{enumerate}

\section{Experimental Results}

The experimental evaluation first introduces the experimental setup, including benchmark circuit sets, scheme parameters, and experimental environment. Three sets of experiments are designed on the benchmark dataset:In the first group, the proposed scheme is compared with rule-based optimization methods and ZX calculus-based optimization methods in terms of the number of two-qubit gates, verifying the effectiveness and superiority of the proposed scheme. In the second group, ablation analyses are conducted by comparing the proposed scheme with variants excluding circuit grouping, the lookahead strategy, and delayed placement optimization, respectively, to validate the effectiveness of different modules in the scheme. In the third group, the impacts of grouping and $k$-step lookahead search on circuit optimization are analyzed, and the influence of $k$ values on circuits is compared to verify the scheme’s scalability.

\subsection{Experimental Setup}
To evaluate the effectiveness of the proposed scheme, comparative experiments are constructed on benchmark circuit sets derived from Nam\cite{namAutomatedOptimizationLarge2018}, which have been used in numerous studies \cite{xuQuartzSuperoptimizationQuantum2022}\cite{hietalaVerifiedOptimizerQuantum2021}\cite{kissingerPyZXLargeScale2020}\cite{staudacherReducing2QuBitGate2023a} for logical circuit gate count optimization. For fair comparison of gate count optimization, the to\_basic\_gates method in pyzx \cite{kissingerPyZXLargeScale2020} is used to convert benchmark circuits into Clifford+T gate sets. The transformed benchmark dataset contains 25 quantum circuits with 5–36 logical qubits and 45–1223 quantum gates. The number of two-qubit gates and total gate count are adopted as evaluation metrics.

The proposed scheme is implemented based on PyZX \cite{kissingerPyZXLargeScale2020} and R2Q \cite{staudacherReducing2QuBitGate2023a}, where PyZX enables ZX diagram conversion or extraction and R2Q obtains matching candidate rules. Parameters are set via trial-and-error and experience as: minimum grouping size $b$=5, lookahead search steps $k$=4, and simulated annealing temperature $T$=300.
\subsection{Comparative Experiments with State-of-the-Art Methods}
To validate the effectiveness and superiority of the proposed scheme, comparisons are conducted with the original circuits, rule-based methods, and ZX calculus-based methods, focusing on the number of two-qubit gates and total gate counts.
\begin{table}[h!t]
	\centering
	\caption{Optimization Comparison of Two-Qubit Gate Count (rule-based)}
	\begin{tabular}{@{}cccccccc@{}}
		
		\toprule
		\textbf{CircuitName}         & \textbf{Qubit} & \textbf{Ori} & \textbf{Nam} & \textbf{VOQC} & \textbf{Qiskit} & \textbf{Quartz} & \textbf{Our} \\ \midrule
		adder\_8                     & 24             & 409          & 291          & 337           & 385             & 407             & 287          \\
		barenco\_tof\_3              & 5              & 24           & 18           & 22            & 24              & 16              & 20           \\
		barenco\_tof\_4              & 7              & 48           & 34           & 44            & 48              & 30              & 37           \\
		barenco\_tof\_5              & 9              & 72           & 50           & 66            & 72              & 48              & 55           \\
		barenco\_tof\_10             & 19             & 192          & 130          & 176           & 192             & 188             & 144          \\
		csla\_mux\_3\_original       & 15             & 80           & 70           & 74            & 71              & 76              & 70           \\
		csum\_mux\_9\_corrected      & 30             & 168          & 140          & 168           & 168             & 166             & 150          \\
		gf2\textasciicircum{}4\_mult & 12             & 99           & 99           & 99            & 99              & 99              & 95           \\
		gf2\textasciicircum{}5\_mult & 15             & 154          & 154          & 154           & 154             & 154             & 149          \\
		gf2\textasciicircum{}6\_mult & 18             & 221          & 221          & 221           & 221             & 221             & 214          \\
		gf2\textasciicircum{}7\_mult & 21             & 300          & 300          & 300           & 300             & 300             & 291          \\
		gf2\textasciicircum{}8\_mult & 24             & 405          & 405          & 405           & 405             & 405             & 393          \\
		gf2\textasciicircum{}9\_mult & 27             & 494          & 494          & 494           & 494             & 494             & 480          \\
		mod5\_4                      & 5              & 28           & 28           & 28            & 28              & 22              & 18           \\
		mod\_mult\_55                & 9              & 48           & 40           & 40            & 48              & 40              & 40           \\
		mod\_red\_21                 & 11             & 105          & 77           & 93            & 105             & 105             & 84           \\
		qcla\_adder\_10              & 36             & 233          & 183          & 207           & 213             & 231             & 186          \\
		qcla\_com\_7                 & 24             & 186          & 132          & 148           & 174             & 182             & 135          \\
		qcla\_mod\_7                 & 26             & 382          & 292          & 338           & 362             & 380             & 305          \\
		rc\_adder\_6                 & 14             & 93           & 71           & 73            & 81              & 73              & 71           \\
		tof\_3                       & 5              & 18           & 14           & 16            & 18              & 14              & 15           \\
		tof\_4                       & 7              & 30           & 22           & 26            & 30              & 22              & 24           \\
		tof\_5                       & 9              & 42           & 30           & 36            & 42              & 30              & 33           \\
		tof\_10                      & 19             & 102          & 70           & 86            & 102             & 90              & 78           \\
		vbe\_adder\_3                & 10             & 70           & 50           & 54            & 62              & 47              & 41           \\\midrule
		reduce            			 &         		  &              &-0.18			& -0.09 		& -0.02			  & -0.12		    &-0.18         \\
		\bottomrule
		
	\end{tabular}
	
	\label{table 1}
\end{table}
\subsubsection{Comparison with Rule-Based Methods}

\begin{figure}[h]
	\centering
	\includegraphics[width=150mm]{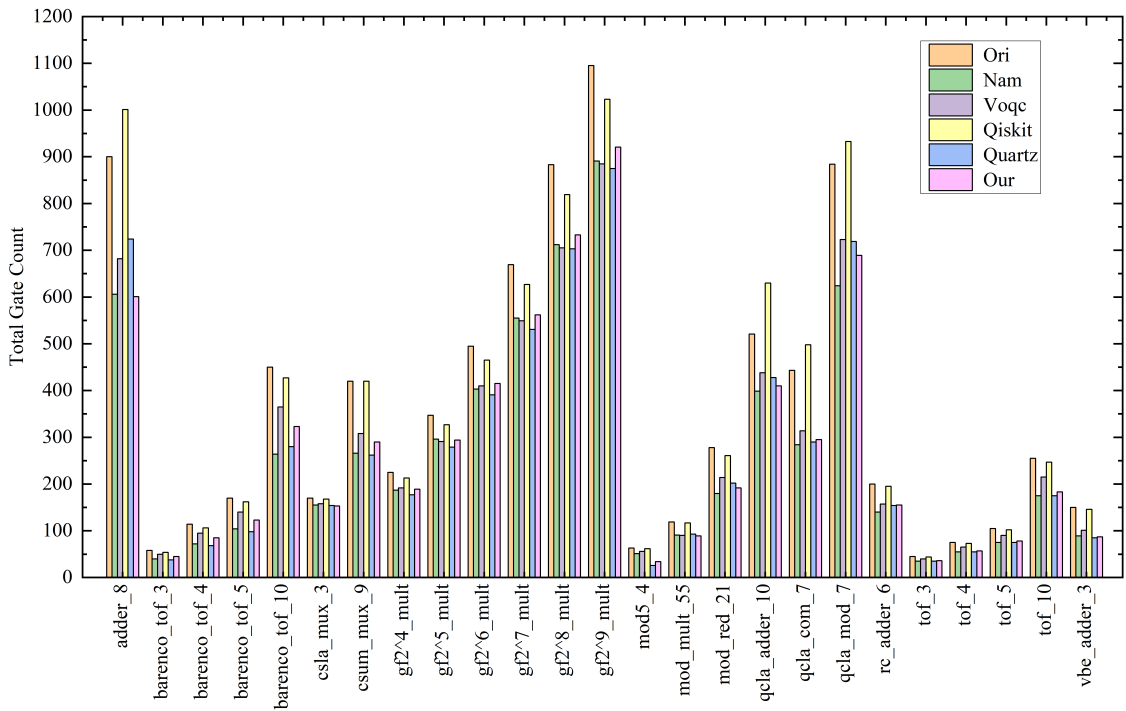}
	\caption{Optimization Comparison of Gate Count (rule-based)}
	\label{fig:figure4-1}
\end{figure}

The proposed scheme is compared with rule-based methods including Nam \cite{namAutomatedOptimizationLarge2018}, VOQC \cite{hietalaVerifiedOptimizerQuantum2021}, Qiskit\cite{javadi-abhariQuantumComputingQiskit2024}, and Quartz \cite{xuQuartzSuperoptimizationQuantum2022}. Performance evaluation in terms of two-qubit gate counts and total gate counts is presented in Table \ref{table 1} and Figure \ref{fig:figure4-1}.

Comparing the proposed scheme with the initial circuits, the two-qubit gate counts of all circuits are reduced. The vbe\_adder\_3 circuit achieves the maximum reduction of 41\%, while the gf2\textasciicircum9\_mult circuit achieves the minimum reduction of 2\%. The average reduction in two-qubit gate count is 18\%, demonstrating the effectiveness of the proposed scheme. Compared with the initial circuits, the average reductions in two-qubit gate counts for Nam, VOQC, Qiskit, Quartz, and the proposed scheme are 18\%, 9\%, 2\%, 12\%, and 18\%, respectively. Our method outperforms the equivalence-class-based Quartz method on circuits such as barenco\_tof, and although the reduction rate is similar to Nam’s, the optimal circuit classes differ. Nam performs better on qcla and tof circuits, while our method successfully optimizes gf circuits, indicating that it can discover structures not detectable by rule-based approaches.However, our method performs less favorably on smaller circuits. This is primarily due to the nature of ZX-calculus-based simplification: although the ZX graph can be simplified effectively, the circuit extraction process may increase the number of two-qubit gates due to the loss of locality in gate placement. As shown in Table \ref{table 1}, \par

Figure \ref{fig:figure4-1} shows that, in terms of total gate count, Nam and Quartz achieve average reductions of 27\% and 28\%, respectively, while our method achieves up to 25\%. This gap arises because the ZX extraction process often introduces additional single- and two-qubit gates. Although phase gadgets within the ZX optimization reduce T gates, they also increase the number of H gates. Therefore, when using ZX calculus for total gate count optimization, the impact of the extraction strategy should be carefully considered.\par

As demonstrated by the experimental data, the proposed approach is effective in reducing the number of two-qubit gates in quantum circuits. However, optimizing two-qubit gates requires precise control over the topological arrangement of gate sequences, and the graph-theoretical simplifications in ZX calculus may disrupt the local structure of gate order, potentially increasing the number of two-qubit gates in the extracted circuit. Moreover, the reduction in two-qubit gates may come at the cost of additional single-qubit gates, which can lead to an overall increase in total gate count. Therefore, the extraction strategy plays a critical role when optimizing quantum circuits using ZX calculus, especially in the context of minimizing total gate count.

\subsubsection{Comparison with ZX Calculus-Based Methods}

To evaluate the effectiveness of our approach, we compare it with two representative ZX-calculus-based optimization methods: PyZX \cite{kissingerPyZXLargeScale2020} and a heuristic method \cite{staudacherReducing2QuBitGate2023a} that integrates random, greedy, and simulated annealing strategies (referred to as R2Q). 

\begin{table}[h]
	\setlength{\tabcolsep}{2pt}
	\centering
	\caption{Optimization Comparison of Two-Qubit Gate Count (ZX calculus-based)}
	\label{key}
	\begin{tabulary}{\textwidth}{@{}ccccccccc@{}}
		\toprule
		\multirow{2}{*}{\textbf{CircuitName}} & \multirow{2}{*}{\textbf{Qubit}} & \multirow{2}{*}{\textbf{Ori}} & \multicolumn{3}{c}{\textbf{ZX-Calculus}}  & \multicolumn{3}{c}{\textbf{Nam+ZX-Calculus}}       \\ \cmidrule(l){4-9} 
		&                                 &                               & \textbf{PyZX} & \textbf{R2Q} & \textbf{Our} & \textbf{PyZX-PP} & \textbf{R2Q-PP} & \textbf{Our-PP} \\ \midrule
		adder\_8                              & 24                              & 409                           & 347         & 295          & 287          & 291            & 256             & 244             \\
		barenco\_tof\_3                       & 5                               & 24                            & 22          & 21           & 20           & 18             & 18              & 18              \\
		barenco\_tof\_4                       & 7                               & 48                            & 44          & 40           & 37           & 34             & 34              & 32              \\
		barenco\_tof\_5                       & 9                               & 72                            & 66          & 57           & 55           & 50             & 48              & 46              \\
		barenco\_tof\_10                      & 19                              & 192                           & 176         & 151          & 144          & 130            & 118             & 116             \\
		csla\_mux\_3\_original                & 15                              & 80                            & 72          & 74           & 70           & 70             & 67              & 66              \\
		csum\_mux\_9\_corrected               & 30                              & 168                           & 168         & 150          & 150          & 140            & 140             & 140             \\
		gf2\textasciicircum{}4\_mult          & 12                              & 99                            & 99          & 101          & 95           & 99             & 98              & 96              \\
		gf2\textasciicircum{}5\_mult          & 15                              & 154                           & 154         & 156          & 149          & 154            & 155             & 147             \\
		gf2\textasciicircum{}6\_mult          & 18                              & 221                           & 221         & 217          & 214          & 221            & 218             & 211             \\
		gf2\textasciicircum{}7\_mult          & 21                              & 300                           & 300         & 299          & 291          & 300            & 292             & 287             \\
		gf2\textasciicircum{}8\_mult          & 24                              & 405                           & 405         & 405          & 393          & 405            & 399             & 389             \\
		gf2\textasciicircum{}9\_mult          & 27                              & 494                           & 494         & 494            & 480          & 494            & 494               & 474             \\
		mod5\_4                               & 5                               & 28                            & 27          & 23           & 18           & 28             & 23              & 22              \\
		mod\_mult\_55                         & 9                               & 48                            & 42          & 40           & 40           & 40             & 41              & 40              \\
		mod\_red\_21                          & 11                              & 105                           & 93          & 85           & 84           & 77             & 76              & 76              \\
		qcla\_adder\_10                       & 36                              & 233                           & 205         & 193          & 186          & 183            & 182             & 179             \\
		qcla\_com\_7                          & 24                              & 186                           & 148         & 138          & 135          & 132            & 132             & 126             \\
		qcla\_mod\_7                          & 26                              & 382                           & 324         & 311          & 305          & 292            & 292               & 282             \\
		rc\_adder\_6                          & 14                              & 93                            & 77          & 71           & 71           & 71             & 71              & 71              \\
		tof\_3                                & 5                               & 18                            & 16          & 14           & 15           & 14             & 14              & 14              \\
		tof\_4                                & 7                               & 30                            & 26          & 24           & 24           & 22             & 22              & 22              \\
		tof\_5                                & 9                               & 42                            & 36          & 40           & 33           & 30             & 30              & 30              \\
		tof\_10                               & 19                              & 102                           & 86          & 78           & 78           & 70             & 70              & 70              \\
		vbe\_adder\_3                         & 10                              & 70                            & 52          & 42           & 41           & 50             & 42              & 43              \\ \midrule
		reduce                                &                                 &                               & -0.09       & -0.15        & -0.18        & -0.18          & -0.20           & -0.22           \\ \bottomrule
	\end{tabulary}
	\label{table 2}
	
\end{table}
As shown in Table \ref{table 2}, our method achieves fewer two-qubit gates than all ZX-based approaches on all circuits except tof\_3. The lower performance on tof\_3 is attributed to its small size—after partitioning, each subcircuit contains fewer two-qubit gates, making it difficult to identify significant optimization opportunities. The PyZX method aims to minimize the ZX diagram by reducing the number of nodes and edges. However, due to limitations in circuit extraction strategies, a minimized ZX graph does not always correspond to a circuit with fewer gates. In contrast, R2Q searches for ZX subgraphs with fewer connections using heuristic techniques. Our method introduces a structured approach that partitions the circuit into multiple subcircuits, applies ZX transformations with filtered candidate rules, and leverages $k$-step lookahead search to select the transformation path that minimizes the number of two-qubit gates. This fine-grained subcircuit optimization combined with guided rule selection enables our method to identify more efficient transformation paths. When using Nam-preprocessed circuits as the input, our Our-PP method achieves an average of 22\% reduction in two-qubit gates, outperforming both PyZX-PP and R2Q-PP. This further confirms the advantage of ZX-calculus-based optimization in identifying non-obvious structures beyond rule-based approaches. Our strategy of local partitioning and anticipatory rule screening contributes to superior optimization outcomes.

\begin{figure}[h]
	\centering
	\includegraphics[width=150mm]{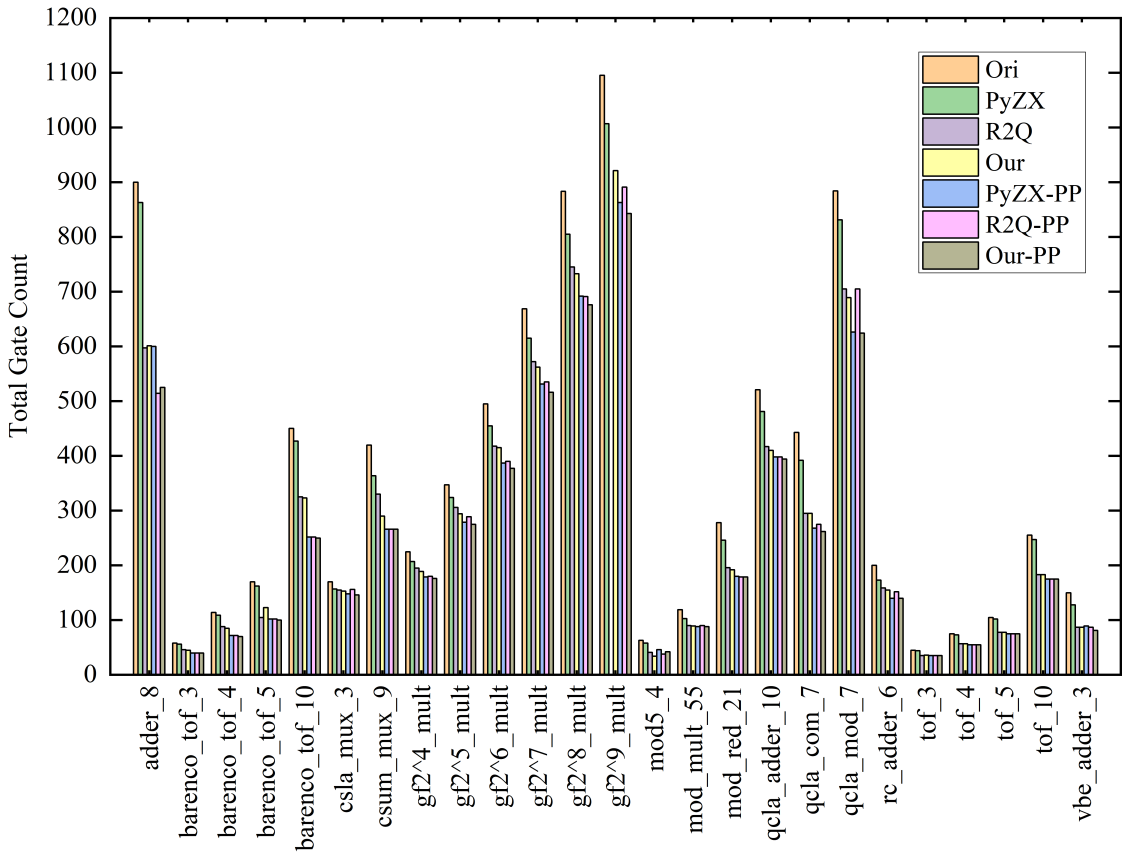}
	\caption{Optimization Comparison of Gate Count (ZX calculus-based)}
	\label{fig:figure4-2}
\end{figure}

Figure \ref{fig:figure4-2} compares the total gate counts. The PyZX method reduces total gates by 8\%, R2Q by 24\%, and our method by 25\%. Despite the fact that ZX-based extraction may introduce additional gates—especially single-qubit gates—the proposed method still outperforms PyZX in terms of total gate count, suggesting that it maintains better circuit structure after transformation. In most cases, our approach also reduces more gates than R2Q. Moreover, when combined with rule-based preprocessing, our method achieves up to 31\% gate count reduction, validating its flexibility and robustness. \par

While our method shows competitive performance, certain circuits such as adder\_8, barenco\_tof\_4, and tof\_3 yield lower total gate counts with R2Q. Notably, for adder\_8 and barenco\_tof\_4, R2Q sacrifices two-qubit gate count in favor of fewer single-qubit gates, indicating a trade-off. In tof\_3, R2Q outperforms our method by one gate in both two-qubit and total counts, suggesting it found a globally optimal structure for this small circuit. This result highlights a limitation of our method: for small-scale circuits, partitioning may hinder optimization. In such cases, applying global ZX transformations without partitioning may lead to better outcomes.

\subsection{Ablation Study}
The proposed optimization framework consists of three main components: circuit partitioning, intra-group optimization, and inter-group optimization. To verify the effectiveness of each module, we conduct ablation experiments by comparing the results with and without each component, including: (1) the partitioning strategy, (2) the $k$-step lookahead rule selection, and (3) the delayed gate placement optimization. The benchmark circuits used are barenco\_tof\_4 (7 qubits, 114 total gates, 48 two-qubit gates) and qcla\_adder\_10 (36 qubits, 521 total gates, 133 two-qubit gates).

\subsubsection{Impact of Grouping Strategy on Circuit Optimization}

\begin{figure}[htbp]
	\centering
	\includegraphics[width=150mm]{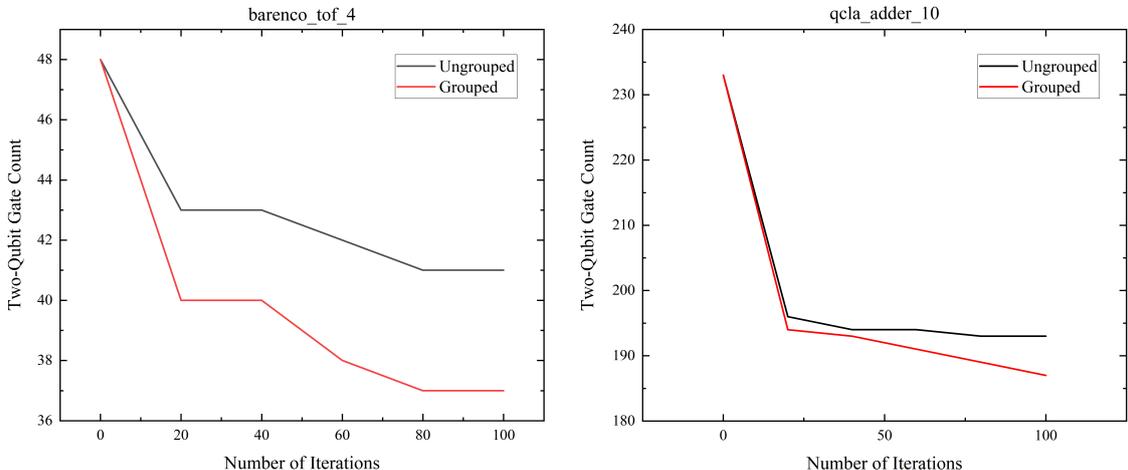}
	\caption{Impact Analysis of Circuit Grouping}
	\label{fig:figure4-3}
\end{figure}

We compare the optimization results with and without the use of partitioning while keeping lookahead and delayed placement enabled. The circuit with partitioning demonstrates a steeper reduction in the number of two-qubit gates over iterations, indicating faster convergence to optimized circuits. Partitioning decomposes the circuit into smaller subcircuits, enlarging the space of ZX diagram rule matches and enabling more opportunities to apply effective local transformations. In contrast, the unpartitioned approach searches for rule applications in the entire circuit, where transformation candidates are limited and less efficient. This validates the effectiveness of the proposed partitioning strategy, as shown in Figure \ref{fig:figure4-3}.

\subsubsection{Effectiveness Verification of Lookahead Strategy}
\begin{figure}[h]
	\centering
	\includegraphics[width=150mm]{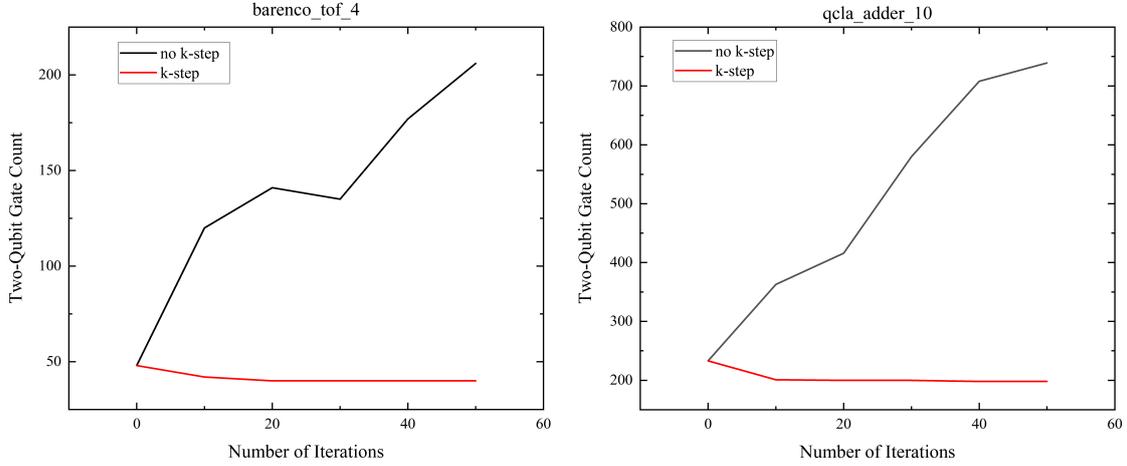}
	\caption{Comparison of the Impact of Lookahead Strategies}
	\label{fig:figure4-4}
\end{figure}

To assess the impact of $k$-step lookahead, we compare optimization results with ($k$-step) and without (no $k$-step) the lookahead mechanism. Both settings utilize partitioning and delayed placement. The use of lookahead results in a consistent decrease in two-qubit gates, while the baseline without lookahead experiences a significant increase. This is because heuristic ZX optimizers apply all matched rules blindly, which, especially after partitioning, may result in additional two-qubit gates in the extracted circuit. In contrast, lookahead-based rule screening filters out transformations that would increase the number of two-qubit gates, even if they simplify the ZX diagram itself. These results demonstrate the advantage of using anticipatory rule selection, as shown in Figure \ref{fig:figure4-4},

\subsubsection{Effectiveness Analysis of Delayed Placement Optimization}

We further compare the results with and without the delayed gate placement optimization ($F_{BO}$), focusing on total gate count (G) and two-qubit gate count (2G). 

\begin{figure}[h]
	\centering
	\includegraphics[width=150mm]{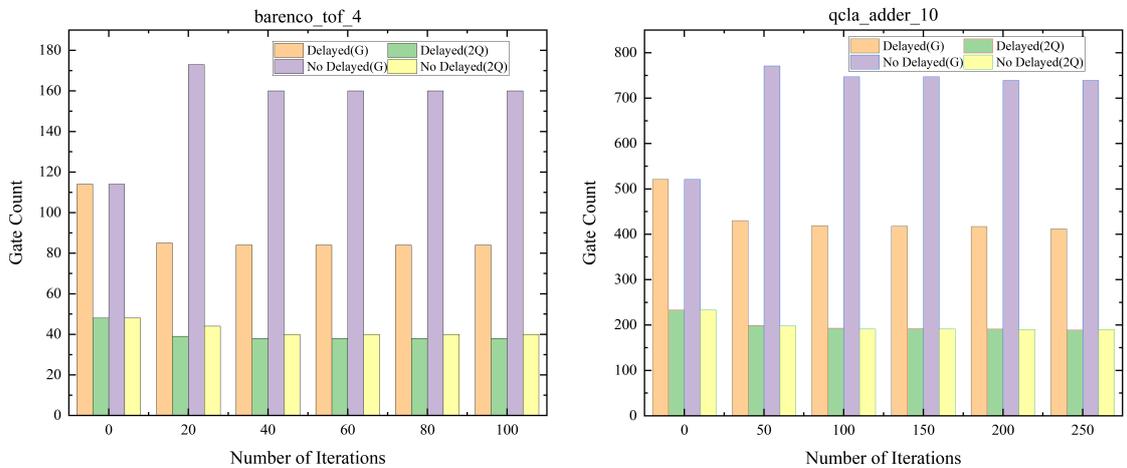}
	\caption{Impact Assessment of Deferred Optimization Strategies}
	\label{fig:figure4-5}
\end{figure}

Both variants produce similar curves in terms of two-qubit gate count. However, the total number of gates increases significantly when delayed placement is not applied. This indicates that delayed placement is particularly effective at reducing redundant single-qubit gates introduced during circuit extraction from ZX diagrams. Since ZX-based optimization focuses on diagram-level structural transformations, the extracted circuits often contain excess single-qubit gates that are not eliminated without gate-level post-processing. The use of $F_{BO}$ effectively suppresses such overhead and confirms its critical role in minimizing total gate count, as shown in Figure \ref{fig:figure4-5},

\subsection{Scalability Analysis}

The performance of the proposed framework is largely influenced by two key parameters: the number of partitions and the lookahead depth $k$. The random partitioning strategy allows flexible decomposition of the original circuit into multiple subcircuits, improving the chance of escaping local optima. Meanwhile, the lookahead depth controls the breadth of rule exploration: larger $k$ enables deeper rule application sequences and potentially yields more optimal transformations. To analyze the effect of lookahead depth, we vary $k$ from 1 to 6 and evaluate the resulting number of two-qubit gates on barenco\_tof\_4 and qcla\_adder\_10.

\begin{figure}[h]
	\centering
	\includegraphics[width=150mm]{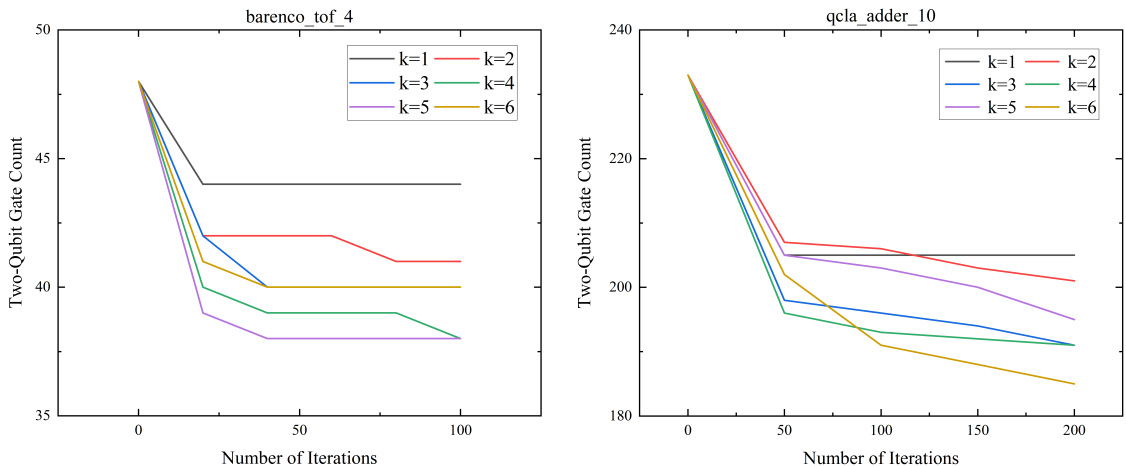}
	\caption{Comparison of Two-Qubit Gate Counts under Different Lookahead Steps}
	\label{fig:figure4-6}
\end{figure}

The optimization follows a common trend: steep improvement at small $k$, followed by diminishing returns as $k$ increases. For barenco\_tof\_4, the worst result occurs at $k$ = 1, while $k$ = 4 and $k$ = 5 yield the best performance. For qcla\_adder\_10, $k$ = 6 gives the best result, while $k$ = 3 and $k$ = 4 are near-optimal. These results indicate that one-step transformations are insufficient for capturing deeper optimization structures. Since the total number of matched rules grows with $k$ × branching factor, increasing $k$ broadens the transformation space and improves the chance of finding better circuits, but at the cost of higher computational overhead.
The proposed scheme's impact on two-qubit gate count stems from the circuit grouping process and $k$-step lookahead search. Randomly partitioning global circuits into sub-circuits avoids local optima, facilitating global optimal partitioning. Varying $k$ in $k$-step lookahead search expands rule-matching scope, enabling optimization by adjusting $k$ to find circuits with minimal two-qubit gates, as shown in Figure \ref{fig:figure4-6},

\section{Conclusion}

This paper proposes a quantum circuit optimization framework based on randomized dynamic partitioning and ZX-calculus-driven rule selection. To reduce circuit complexity, we first introduce a dynamic partitioning strategy that decomposes the circuit into subcircuits using randomized grouping. For intra-group optimization, we develop a $k$-step lookahead ZX-transformation method, which converts each subcircuit into a ZX diagram and filters transformation candidates over $k$ steps to extract an equivalent circuit with fewer two-qubit gates. Subsequently, we introduce a delayed gate placement scheme to recombine subcircuits into a global circuit, and apply post-processing optimizations to minimize the number of redundant single-qubit gates, particularly H gates. The entire process is embedded within a simulated annealing loop to iteratively refine partitioning decisions and search for global optima.

Experimental results on benchmark circuits and randomly generated instances demonstrate the effectiveness of the proposed scheme. On average, our method reduces the number of two-qubit gates by 8\% compared to delay-placement-only optimization, and by 3\% compared to ZX-based heuristic methods. In some cases, the proposed method further improves the output of the Nam optimizer by up to 15\%, validating the value of the integrated partitioned lookahead strategy.

Currently, the optimization focuses on minimizing the number of two-qubit gates. However, quantum circuit quality also depends on other metrics such as total gate count, fidelity, and circuit depth. While total gate count can be incorporated via cost function adjustment, fidelity and depth are also influenced by the physical hardware layout and the circuit mapping strategy. Since the ZX extraction process introduces nontrivial effects on circuit structure, directly assessing depth or fidelity from ZX diagrams remains challenging. These aspects are left for future investigation.

Future work will focus on the following directions:
\begin{enumerate}[$\bullet$]
	\item Multi-objective optimization, including depth and fidelity, with consideration of device-aware circuit mapping and improved ZX extraction heuristics.
	\item Scalable rule search algorithms, as the current approach scales poorly with increasing circuit size due to the combinatorial complexity of rule matching.
	\item Learning-based optimization frameworks, such as reinforcement learning, to dynamically guide transformation selection and grouping strategies.
	\item Connectivity-aware optimization, targeting reductions in post-mapping two-qubit gate count to enhance physical implementation performance.
\end{enumerate}

Overall, the proposed approach bridges structural ZX simplification and gate-level cost-aware optimization, offering a promising direction for scalable and effective quantum circuit synthesis.

\bibliography{ref}
\bibliographystyle{quantum}

\end{document}